\documentclass[lettersize,journal]{IEEEtran}
\usepackage{caption}
\usepackage{amsmath,amsfonts}
\usepackage{algorithmic}
\usepackage{algorithm}
\usepackage{array}
\usepackage{textcomp}
\usepackage{stfloats}
\usepackage{multicol}
\usepackage{placeins}
\usepackage{url}
\usepackage{siunitx}
\usepackage{mathtools}
\usepackage{verbatim}
\usepackage{graphicx}
\usepackage{cite}
\usepackage{dsfont}
\usepackage{makecell}
\usepackage{color}
\usepackage{float}
\usepackage{makecell}
\usepackage{tabularx,colortbl}
\usepackage{ifthen}
\usepackage{caption}
\usepackage{diagbox}
\usepackage{nicematrix}
\usepackage{subcaption}
\usepackage{svg}

\newcommand{\NRWG}{N}
\newcommand{\NRWGL}{N^l}
\newcommand{\NRWGS}{N^s}
\newcommand{\NPC}{M}
\newcommand{\NPCO}{M^o}

\begin{document}

\title{Broadband Stable Calderón-Preconditioned\\ Vector-Potential-Only Integral Equations \\for PEC Scattering}

\author{Paul Olyslager, Hendrik Rogier and Kristof Cools

\thanks{This project has received funding from the European Re-
search Council (ERC) under the European Union’s Horizon
2020 research and innovation programme (Grant agreement
No. 101001847).}
\thanks{P. O. Author, Dept. of Information Technology,
Ghent University, Ghent, Belgium (Paul.Olyslager@ugent.be)}
\thanks{H. R. Author, Dept. of Information Technology,
Ghent University, Ghent, Belgium (Hendrik.Rogier@ugent.be)}
\thanks{K. C. Author, Dept. of Information Technology,
Ghent University, Ghent, Belgium (Kristof.Cools@ugent.be)}

}

\IEEEpubid{}

\maketitle

\begin{abstract}
We propose a stabilized first-kind, Calderón preconditioned, vector-potential-only integral equation, for scattering at perfect electrical conductors. The method yields accurate low-frequency results for both the vector and the scalar potential, where the scalar potential is computed using the Lorenz gauge in the post-processing. Both the near and far fields can be accurately computed at arbitrarily low frequencies. This is achieved in a post-processing step and is independent of machine and quadrature precision. First, the low-frequency scaling of the degrees of freedom, and the near and far fields, are studied. Next, stabilized methods are derived based on the low-frequency scaling. The tangential trace of the curl of the vector-potential is stabilized leveraging quasi-Helmholtz projectors, while the normal trace of the vector-potential is stabilized with a projector based on the static scalar potential. Numerical experiments are presented showing that the stabilized method shows accurate results even for very low frequencies.

\end{abstract}
\section{Introduction}
In multi-physics modeling, it is often essential to directly access the vector and scalar potential, rather than the electric and magnetic fields. Among others, this is specifically the case when the Maxwell system is coupled to the Schrödinger or Dirac equation \cite{EMQMFDTD,EMQMFinite_element}. In fact, it is known that some quantum-electrodynamical phenomena such as the Aharonov-Bohm effect, can only be understood in terms of these potentials \cite{vico_decoupled_2016}. Therefore, we are interested in computing the potentials directly instead of the electromagnetic fields. Often, these multi-physics scenarios are modelled using finite-difference and finite-element methods. Unfortunately, approaches based on these methods are susceptible to spatial dispersion errors, spurious reflections at the boundary of the simulation domain, and inaccuracies in describing some of the small geometric details that may appear in the system geometry. Moreover, these methods require the introduction of degrees of freedom in the entire simulation domain.
For these reasons, it is therefore of interest to have access to a boundary element method (BEM) into the EM-QM coupling framework. Notably, the EM-QM interaction is inherently non-linear and demands time-domain electromagnetic solvers capable of maintaining accuracy across a wide frequency spectrum and for long simulation times. In this contribution, we take an initial step toward the development of such a multi-physics solver by proposing a frequency-domain formulation that remains stable over a broad range of frequencies. Although the present work does not yet resolve all the aforementioned challenges, it provides a foundation for future developments toward a comprehensive and robust EM–QM coupling framework.
    
The design of time-stable vector-potential boundary element methods is closely linked to the development of low-frequency-stable techniques in the frequency-domain \cite{TD-FD_relation}. It is therefore essential to first fully develop a low-frequency-stable vector-potential integral equation method up to arbitrary low frequencies. In fact, so-called DC instabilities in time-domain solvers can be directly traced back to low-frequency breakdown of the corresponding frequency-domain method \cite{chienlftd}.

An additional requirement on the numerical stability of the solver is that it should be possible to derive the solution for the scalar potential from the vector-potential by applying the Lorenz gauge without loss of physically relevant information, thereby eliminating the need to solve an additional scalar Helmholtz equation. This implies that the proposed stabilization procedures are also required to provide accurate electric near-field information. The fields and potentials obtained from the method  should be accurate in both the near and far field up to some controllable tolerance, related to the mesh size and independent of the frequency. To obtain the desired accuracy at low frequencies, a Helmholtz splitting and rescaling method based on projectors, following the cue of similar methods used in the modelling of the fields is followed \cite{andriulli_well-conditioned_2013}. Such a projector based rescaling does not significantly degrade the condition number compared to an explicit loop-star decomposition, which has a detrimental effect on the dense-grid properties of the boundary element method. Besides, the explicit loop-star decomposition cannot be easily generalized to work for scatterer geometries exhibiting a general topology. \cite{QHP_VS_LS}.

The number of iterations of the generalized-minimal-residual (GMRES) iterative Krylov solver should not increase when the mesh is refined. This is achieved by applying a Calderón preconditioner.

 The low-frequency stabilization of both the electric field integral equation (EFIE) \cite{QHP_VS_LS,andriulli_well-conditioned_2013,Excitation-Aware-Bernd} and the magnetic field integral equation (MFIE) \cite{MFIE_LOW_FREQ,MFIE_Low_Freq_2} has been extensively studied. These stabilization techniques are typically based on either an explicit loop-star or loop-tree decomposition, or on so-called quasi-Helmholtz projectors. Quasi-Helmholtz projectors project the boundary element space on subspaces of solenoidal and approximately irrotational subspaces. The use of quasi-Helmholtz projectors enables the solution of the low-frequency breakdown by rescaling the solenoidal and non-solenoidal components of the trial and testing vectors with different powers of the frequency. When used in conjunction with Calderón preconditioning they can give rise to methods that are accurate and efficient in both the low-frequency and dense-grid regimes. 
    A unified stabilization approach for the EFIE, applicable to arbitrary incident fields and based on quasi-Helmholtz projectors, was proposed in \cite{Excitation-Aware-Bernd}. 
    
   The mathematical foundation of the vector-potential integral equation is discussed in \cite{claeys_first-kind_nodate,schulz_coupled_2022,vico_decoupled_2016}. It is shown that the vector-potential integral equation can be made low-frequency stable, meaning that the solution remains bounded when the incident wave is bounded with respect to the frequency. This is done by introducing an extra degree of freedom (DOF), $V_A$, along with a strong charge-neutrality condition \cite{vico_decoupled_2016}, $\rho_{\boldsymbol{A}}=0$, where $\rho_{\boldsymbol{A}}$ is the integral of the normal trace of the vector potential. However, this analytical stabilization does not account for numerical errors stemming from the application of quadrature rules, iterative inversion of matrices, and loss of significant digits due to round-off errors. In this paper, we further analyze both the formulation that includes $V_A$ as an unknown together with the charge-neutrality condition, and the formulation that is derived under the assumption that $V_A=0$ \cite{VPIE_C_chew}. We then propose strategies to achieve accurate results at arbitrarily low frequencies, in both the near- and far fields, for the vector potential, scalar potential, and  electric and magnetic field, even in the presence of unavoidable numerical errors. The vector-potential formulation is used to model lossy conductors in \cite{Triv_prec_los_cond_pot, sharma_electromagnetic_2022} because of its stability across a broad frequency range.
   
   In \cite{hawkins_analytic_2023}, a Calderón preconditioned CFIE-like formulation for the vector-potential is proposed that is shown to work at moderate frequencies, but no further low-frequency stabilization is applied to investigate the behavior beyond machine precision. The Calderón preconditioner in this paper is discretized on the primal mesh with Rao-Wilton-Glisson basis functions and the wave number in the preconditioner is dependent on the local geometry of the mesh. In this paper, we construct a Calder\'on preconditioner based on a discretization of the regularizer on the dual mesh using Buffa-Christiansen elements, guaranteeing a bounded condition number independent of the meshing parameter $h$. In most of the aforementioned studies,  the vector and scalar potential integral equations are treated jointly, in what is referred to as the decoupled-potential integral equations. A low-frequency stable vector-potential integral equation is proposed in \cite{gur_low-frequency_2017}, where stabilization is achieved by solving the scalar Helmholtz equation in a post-processing step. Before translating such a method to the time-domain, it might be beneficial to start from a method that is inherently low-frequency stable for all degrees of freedom before the post-processing is started. 
   Similarly, in \cite{chen_low-frequency_2022}, the low-frequency scaling of the vector-potential integral equation, without the auxiliary unknown $V_A$, is analyzed by applying a loop-star decomposition on the vectorial degrees of freedom. However, stabilization of the scalar degrees of freedom is not addressed, which is a necessary condition to compute accurate far-field results at frequencies beyond machine precision. 
   In \cite{GlobalMultiCaldPrec} the authors looked into Calderón preconditioners for the vector-potential integral equation for scattering at piecewise homogeneous media at moderate frequencies.
   
   In this contribution, both formulations, the one including the additional unknown $V_A$ with the stronger charge-neutrality condition $\rho_{\boldsymbol{A}}=0$, and the one with $V_A=0$, are Calderón preconditioned and numerically stabilized using a projector-based strategy. The resulting fields and potentials are computed in both the near and far field. This demonstrates that these quantities can be accurately obtained across all frequencies. In particular, the lowest frequency at which the method can be relied on is limited neither by quadrature nor by machine precision.
   
The paper is organized as follows. In Section \ref{sec:mathematical_formulation}, the mathematical background of both formulations considered in this work is revisited, along with a discussion of their equivalence. Section \ref{sec:integral_equations} introduces the corresponding integral equation formulations. The discretization of these equations is described in Section \ref{sec:Discretization}. Section \ref{sec:low_freq_stab} elucidates the origin of numerical instability in the extremely low-frequency regime and outlines the projector-based stabilization strategy to address low-frequency breakdown. Finally, in Section~\ref{sec:results} numerical results are presented that demonstrate that the vector-potential and all secondary quantities of physical interest can be stably retrieved at arbitrarily low frequencies without being limited by quadrature precision, Krylov solver convergence criteria, or round-off errors.

\section{Mathematical Formulation}\label{sec:mathematical_formulation}
A single, but possibly multiply connected 3D PEC domain $\Omega$ with boundary $\Gamma$, for example the one shown in Fig. \ref{fig:holes_mesh}, is embedded in the free space $\Omega_0$. The surface $\Gamma$ is oriented with an outward-pointing normal vector $\boldsymbol{n}$.
 The scalar and tangential traces are defined as continuous extensions to the appropriate Sobolev spaces of the expressions, valid for smooth functions:
\begin{align}
    \tau\left[b\right]\left(\boldsymbol{r}\right) &\vcentcolon= \lim_{\boldsymbol{r}'\in\Omega_0\to\boldsymbol{r}}b\left(\boldsymbol{r}'\right) \quad \boldsymbol{r}\in\Gamma,\\
    \gamma\left[\boldsymbol{b}\right]\left(\boldsymbol{r}\right) &\vcentcolon= \lim_{\boldsymbol{r}'\in\Omega_0\to\boldsymbol{r}}\boldsymbol{n}\times\left[\boldsymbol{b}\left(\boldsymbol{r}'\right)\times\boldsymbol{n}\right] \quad \boldsymbol{r}\in\Gamma,
\end{align}
with $b\in\mathcal{C}^\infty\left(\Omega\right)$ and $\boldsymbol{b}\in\left[\mathcal{C}^\infty\left(\Omega\right)\right]^3$ smooth (vector) functions.
All fields are assumed to vary harmonic-in-time with frequency $\omega$. The time dependency $e^{\jmath\omega t}$ is suppressed throughout this paper. The wavenumber $\kappa$ is given by $\omega\sqrt{\epsilon_0\mu_0}$, with $\epsilon_0$ and $\mu_0$ the free space permittivity and permeability.\\ The scalar and vector duality pairings are defined as 
\begin{equation}\label{eq:dualpair}
    \left<a,b\right>_\Gamma \vcentcolon= \int_\Gamma  a\cdot b \quad dA.
\end{equation}
The (scalar, vector) potential pair $(\phi$, $\boldsymbol{A}$) in $\Omega_0$ is the sum of the incident $(\phi^\text{in}, \boldsymbol{A}^\text{in})$ and the scattered (vector) potential $(\phi^\text{sc}, \boldsymbol{A}^\text{sc})$.  
The scattered and incident vector and scalar potentials should be related by the Lorenz gauge
\begin{equation}\label{eq:Lorenz}
    \nabla\cdot\boldsymbol{A}^{sc/in} = -\jmath\epsilon_0\mu_0\omega\phi^{sc/in}.
\end{equation}
They are also the solution of the following set of equations together with the Sommerfeld radiation conditions defined in \cite{vico_decoupled_2016}:
\begin{align}\label{eq:scalarHelmholtz}
    \nabla^2 &\phi^{sc} + \kappa^2\phi^{sc} = 0\\
\label{eq:bcphi}
     V_\phi &\vcentcolon= \tau\left[\phi^{sc}\right] + \tau\left[\phi^{in}\right], \quad V_\phi \in \mathbb{C} \\
\label{eq:chargescalar}
    \rho_\phi &\vcentcolon=\int_\Gamma \partial_{\boldsymbol{n}}\phi\quad dA, \quad \rho_\phi \in \mathbb{C} \\
\label{eq:Helmholtzhodge}
    \nabla&\times\nabla\times\boldsymbol{A}^{sc} - \nabla\nabla\cdot\boldsymbol{A}^{sc} - \kappa^2\boldsymbol{A}^{sc} = 0\\
\label{eq:bcnA}
    0 &= \gamma\left[\boldsymbol{A}^{sc}\right] + \gamma\left[\boldsymbol{A}^{in}\right]\\
\label{eq:bcdA}
   V_{\boldsymbol{A}}&\vcentcolon= \tau\left[\nabla\cdot\boldsymbol{A}^{sc}\right] + \tau\left[\nabla\cdot\boldsymbol{A}^{in}\right] , \quad V_{\boldsymbol{A}} \in \mathbb{C} \\
\label{eq:chargevector}
    \rho_{\boldsymbol{A}} &\vcentcolon = \int_\Gamma \boldsymbol{n}\cdot\boldsymbol{A} \quad dA, \quad \rho_{\boldsymbol{A}} \in \mathbb{C}  ,
\end{align}
where \eqref{eq:bcphi} and \eqref{eq:bcdA} indicate that the trace is constant on the surface. The parameters $V_\phi$, $V_{\boldsymbol{A}}$ $\rho_\phi$ and $\rho_{\boldsymbol{A}}$ cannot be chosen freely. We will see in the next part that they are related and that, at $\omega \neq 0$ specifying one of them is enough to obtain a unique solution.
In the next paragraph we discuss how $V_{\boldsymbol{A}}$ is related, to $V_\phi$, $\rho_\phi$ and $\rho_{\boldsymbol{A}}$, depending on frequency.

For any set of potentials that satisfy equations \eqref{eq:scalarHelmholtz}-\eqref{eq:chargevector} and the Lorenz gauge, $V_\phi$ and $V_{\boldsymbol{A}}$ are related by the following expression
\begin{equation}\label{eq:Vrelation}
    V_{\boldsymbol{A}} = -\jmath\epsilon_0\mu_0\omega V_\phi.
\end{equation} 
In the next paragraph we prove the corresponding relation between $\rho_{\boldsymbol{A}}$ and $\rho_\phi$, and show that they can be nonzero. Demanding that they both be zero is a stronger charge-neutrality condition than the physical charge-neutrality condition stated below in \eqref{eq:charge1}.
The pair $\left(\phi,\boldsymbol{A}\right)$ is a solution of equations \eqref{eq:scalarHelmholtz}-\eqref{eq:chargevector} and the Sommerfeld radiation condition, with incident fields $(\phi^\text{in},\boldsymbol{A}^\text{in})$ and $\rho_\phi = \rho_{\boldsymbol{A}} = 0$ for some $V_{\boldsymbol{A}}$ and $V_\phi$. The total charge $\rho$ on the object $\Omega$ can be expressed as
\begin{equation}\label{eq:charge1}
    \rho \vcentcolon= \epsilon_0\int_\Gamma \boldsymbol{n}\cdot\boldsymbol{E} \quad dA = -\jmath\omega\epsilon_0\rho_{\boldsymbol{A}} - \epsilon_0\rho_\phi = 0
\end{equation}
and is equal to zero. We introduce the scalar field $\chi$, which is the unique solution of the following equations and the Sommerfeld radiation condition:
\begin{equation}
    \nabla^2\chi+\kappa^2\chi = 0
\end{equation}
\begin{equation}
    \tau\left[\chi\right] = -\frac{\jmath V_\phi}{\omega}.
\end{equation}
The fields $\left(\phi',\boldsymbol{A}'\right) = \left(\phi -\jmath\omega\chi,\boldsymbol{A}+\nabla\chi\right)$ fulfill the equations \eqref{eq:scalarHelmholtz}-\eqref{eq:chargevector} with $V_\phi=V_{\boldsymbol{A}} = 0$. A gauge transformation does not affect the electric or magnetic field, and thus the total charge $\rho$ obtained from \eqref{eq:charge1} is conserved and equal to zero. However, the contributions $\rho_\phi$ and $\rho_{\boldsymbol{A}}$ to the total charge can be nonzero, which concludes the proof. Moreover, equation \eqref{eq:charge1} yields the relation between $\rho_\phi$ and $ \rho_{\boldsymbol{A}}$.

Equations \eqref{eq:Helmholtzhodge}-\eqref{eq:bcdA} together with the Sommerfeld radiation condition, posses a unique solution at $\omega \neq 0$ \cite{vico_decoupled_2016}. Therefore, there is a unique bijective map between the potential parameter $V_{\boldsymbol{A}}$ and $\rho_{\boldsymbol{A}}$. This can be proven by contradiction. Assume that $V_{\boldsymbol{A},1}$ and $V_{\boldsymbol{A},2}$ both map to $\rho_{\boldsymbol{A}}$, then by subtracting both solutions, a solution is found that maps $V_{\boldsymbol{A},1}-V_{\boldsymbol{A},2}$ to $\rho_{\boldsymbol{A}}=0$, which contradicts Theorem 3.10 from \cite{vico_decoupled_2016}.
Therefore, there is a unique relationship between $V_{\boldsymbol{A}}$ and $\rho_{\boldsymbol{A}}$.

The breakdown at $\omega = 0$ can be understood from \eqref{eq:Vrelation}. When the frequency is zero, and $V_\phi$ remains bounded, the potential parameter satisfies $V_{\boldsymbol{A}} = 0$ independent of $\rho_{\boldsymbol{A}}$. As a result, the charge parameter $\rho_{\boldsymbol{A}}$ is no longer uniquely determined by $V_{\boldsymbol{A}}$, and must instead be specified explicitly as an arbitrary but fixed constant.
This method, proposed in \cite{vico_decoupled_2016}, is referred to as VPIE-V in the remainder of this paper. The fixed value chosen will be $\rho_{\boldsymbol{A}} = 0$.

We propose a second low-frequency stabilization strategy applied directly to \eqref{eq:Helmholtzhodge}-\eqref{eq:bcdA}, in which the potential parameter $V_{\boldsymbol{A}}$ is set to zero. Both contributions to the right-hand side of equation \eqref{eq:bcdA} scale linearly with the frequency, which is explained by the Lorenz gauge \eqref{eq:Lorenz}. This boundary condition is therefore rescaled with $\frac{1}{\omega}$. This operation amounts to a simple rescaling and does not alter the solution at nonzero frequencies. After rescaling, the resulting equation is proportional to \eqref{eq:bcphi} because of the Lorenz gauge. The quantity $\frac{V_{\boldsymbol{A}}}{\omega} = -j\epsilon_0\mu_0 V_\phi = 0$ is associated with a uniquely determined charge density $\rho_\phi$, since the scalar potential problem admits a unique solution at all frequencies \cite{vico_decoupled_2016}. Using \eqref{eq:charge1}, it follows that the charge parameter $\rho_{\boldsymbol{A}}$ scales as $\omega^{-1}$ in the low-frequency limit. Numerically, this behavior is handled by decomposing the $\boldsymbol{n}\cdot\boldsymbol{A}$ term into two components: one lying in the subspace that contributes to $\rho_{\boldsymbol{A}}$, which is rescaled by $\omega$, and its orthogonal complement, which does not contribute to $\rho_{\boldsymbol{A}}$ and is therefore left unscaled.

The integral equations are developed in the next section. Subsequently, the discretization is introduced and stabilization criteria based on the discrete matrices are proposed, which, in the end, for the VPIE-C (the method where $\rho_{\boldsymbol{A}}$), corresponds to the stabilization method described above.

\section{Integral Equations}\label{sec:integral_equations}
The single- and double-layer potential operators are defined as
\begin{align}
    \mathcal{S}_\kappa\left[b\right]\left(\boldsymbol{r}\right) &\vcentcolon= \int_\Gamma G_\kappa\left(\boldsymbol{r},\boldsymbol{r'}\right)b\left(\boldsymbol{r}'\right)d\boldsymbol{A}'
\\
    \mathcal{N}_\kappa\left[b\right]\left(\boldsymbol{r}\right) &\vcentcolon= \int_\Gamma \boldsymbol{n'}\cdot\nabla G_\kappa\left(\boldsymbol{r},\boldsymbol{r'}\right)b\left(\boldsymbol{r}'\right)d\boldsymbol{A}',
\end{align}
where $b$ can be either a scalar or vector function defined on $\Gamma$ and $G_\kappa$ is the three-dimensional Helmholtz Green function, defined as
\begin{equation}
\label{eq:greenfunction}
    G_\kappa\left(\boldsymbol{r},\boldsymbol{r'}\right) \vcentcolon=\frac{e^{-\jmath\kappa\left|\boldsymbol{r}-\boldsymbol{r'}\right|}}{4\pi\left|\boldsymbol{r}-\boldsymbol{r'}\right|} .
\end{equation}
Assume $\boldsymbol{A}^{sc}$ is the solution of \eqref{eq:Helmholtzhodge}-\eqref{eq:chargevector}, then we can write the scattered vector potential in the surrounding free-space region $\Omega_0$ by virtue of the representation theorem as a function of the traces of the total field as \cite{hawkins_analytic_2023}:
    \begin{equation}
        \label{eq:Asc}
    \boldsymbol{A}^\text{sc} \vcentcolon= \mathcal{S}_\kappa\left[\boldsymbol{v}\right] - \nabla\mathcal{S}_\kappa\left[w\right] - \mathcal{S}_\kappa\left[\boldsymbol{n}V_{\boldsymbol{A}}\right],
    \end{equation}

where $\boldsymbol{v}$ and $w$ are unknown `currents' and `charges' with support on $\Gamma$ and $V_{\boldsymbol{A}}$ the potential parameter. They are given by
\begin{align}
    \boldsymbol{v} &\vcentcolon= \boldsymbol{n}\times\nabla\times\boldsymbol{A}|_{\Gamma^+}
\\
    w &\vcentcolon= \boldsymbol{n}\cdot\boldsymbol{A} |_{\Gamma^+},
\end{align}
where $\Gamma^+$ indicates that the limit is taken from within $\Omega_0$.

The trace operators are combined with the potential operators into boundary integral operators, defined as

\begin{align}
    S_\kappa&\vcentcolon= \tau\circ\mathcal{S}_\kappa\\
    N_\kappa &\vcentcolon= \text{P.v.}\left(\tau \circ \mathcal{N}_\kappa\right)
\\
    S_\kappa &\vcentcolon= \gamma\circ\mathcal{S}_\kappa,
\end{align}
with P.v. the principal value.
Applying the trace operators and boundary conditions from \eqref{eq:bcnA} and \eqref{eq:bcdA} to \eqref{eq:Asc}, reveals that $\boldsymbol{v}$ and $w$ are the solutions to the following system of first-kind boundary integral equations:

\begin{align}
\label{eq:VPIE-1}\begin{split}
   \gamma\left[\boldsymbol{A}^\text{in}\right] &=  -S_\kappa\left[\boldsymbol{v}\right] +  \nabla S_\kappa\left[w\right] + {S}\left[\boldsymbol{n}V_{\boldsymbol{A}}\right] \end{split}
\\
\label{eq:VPIE-2}
\begin{split}
       \tau\left[\nabla\cdot\boldsymbol{A}^\text{in}\right] &= -\nabla\cdot{S}_\kappa\left[\boldsymbol{v}\right] -  \kappa^2{S}_\kappa\left[w\right] + N_\kappa\left[V_{\boldsymbol{A}}\right]+ \frac{V_{\boldsymbol{A}}}{2}.
       \end{split}
\end{align}
The formulation consisting of the vector-potential integral equations \eqref{eq:VPIE-1} and \eqref{eq:VPIE-2}, together with equation \eqref{eq:chargevector}, is called the VPIE-V. The formulation with the integral equations \eqref{eq:VPIE-1} and \eqref{eq:VPIE-2} in which the potential parameter $V_{\boldsymbol{A}}$ is set to zero and in which equation \eqref{eq:chargevector} is not included, is denoted by VPIE-C. Note that we will only look at low frequencies, lower than the first internal resonance frequency of the object, such that these systems of boundary operators allow for a unique solution.

\section{Discretization}\label{sec:Discretization}
The surface $\Gamma$ is approximated with a triangular mesh. The scalar trace $w$ is discretized with a piecewise constant (PC) $\left\{f_i\right\}_i$, $i = 1, .. ,\NPC$ space, where $\NPC$ is the dimension of the PC space. The tangential trace $\boldsymbol{v}$ is discretized in a Rao-Wilton-Glisson (RWG) $\left\{\boldsymbol{f}_i\right\}_i$, $i = 1,...,\NRWG$ space, with $\NRWG$ the dimension of the RWG space \cite{rao_electromagnetic_1982}. The same spaces are used to test the system, yielding the following discrete matrix for the VPIE-V, with v and w the coefficient vectors
\begin{equation}\label{eq:discvpie-v}
\begin{bmatrix}
\text{Z}^{11} & \text{Z}^{12} & \text{Z}^{13}\\
\text{Z}^{21} & \text{Z}^{22} & \text{Z}^{23}\\
0_{1\times \NRWG}&\text{Z}^{32}&0_{1\times 1}
    \end{bmatrix}
    \begin{bmatrix}
    \text{v}\\
    \text{w}\\
    \text{V}_{\boldsymbol{A}}
    \end{bmatrix}
     =
     \begin{bmatrix}
     \text{y}\\
     \text{z}\\
     0
     \end{bmatrix},
\end{equation}
with 
    \begin{align*}
    \text{Z}^{11}_{i,j} &\vcentcolon= -\left<\boldsymbol{f}_i,S_\kappa\left[\boldsymbol{f}_j\right]\right>_\Gamma 
    & \text{Z}^{12}_{i,j} &\vcentcolon= -\left<\nabla\cdot\boldsymbol{f}_i, S_\kappa\left[f_j\right]\right>_\Gamma \\
    \text{Z}^{21}_{i,j} &\vcentcolon= \text{Z}^{12}_{j,i} 
    & \text{Z}^{22}_{i,j} &\vcentcolon= -\kappa^2\left<f_i,S_\kappa\left[f_j\right]\right>_\Gamma \\
    \text{Z}^{13}_{i,1} &\vcentcolon= \left<\boldsymbol{f}_i,S_\kappa\left[\boldsymbol{n}\right]\right>_\Gamma 
    & \text{Z}^{23}_{i,1} &\vcentcolon= \left<f_i,N_\kappa\left[1\right] + \frac{1}{2}\right>_\Gamma \\
    \text{Z}^{32}_{1,j} &\vcentcolon= \left<1,f_j\right>_\Gamma 
    & \text{y}_i &\vcentcolon= \left<\boldsymbol{f}_i,\boldsymbol{A}^\text{in}\right>_\Gamma \\
    \text{z}_i &\vcentcolon= \left<f_i,\nabla\cdot\boldsymbol{A}^\text{in}\right>_\Gamma.
\end{align*}

The discrete equation of the VPIE-C is given by
\begin{equation}\label{eq:disc-vpie-c}
\begin{bmatrix}
\text{Z}^{11} & \text{Z}^{12} \\
\text{Z}^{21} & \text{Z}^{22}\\

    \end{bmatrix}
    \begin{bmatrix}
    \text{v}\\
    \text{w}
    \end{bmatrix}
     =
     \begin{bmatrix}
     \text{y}\\
     \text{z}
     \end{bmatrix}.
\end{equation}
The discretized matrix of the VPIE-C method is hereafter denoted as $\text{Z}_\text{C}$ and the discretized matrix of the VPIE-V method as $\text{Z}_\text{V}$.
The incident wave considered in this paper is given by \cite{vico_decoupled_2016}
\begin{equation}\label{eq:incident}
    \boldsymbol{A}^\text{in}\left(x,y,z\right) = x\sqrt{\epsilon_0\mu_0}e^{-i\kappa z}\boldsymbol{1_z}.
\end{equation}
\section{Low-Frequency Stabilization}\label{sec:low_freq_stab}
\subsection{Basis Decomposition}
The RWG and PC spaces are decomposed into orthogonal subspaces to analyze the low-frequency behavior of the corresponding degrees of freedom. Projection operators are then employed to map the solution vector to the respective subspace and to rescale with some power of $\kappa$, since an explicit basis decomposition can severely deteriorate the condition number of the full system \cite{andriulli_cond_loop_star}. In the following paragraphs, however, we first examine the low-frequency behavior in the decomposed basis in order to motivate and derive the proposed stabilization schemes.
The RWG space is separated into a loop-star basis, denoted as $\{\boldsymbol{f}^l_i\}_i\oplus\left\{\boldsymbol{f}^s_j\right\}_j$, where both, the global and local loops are included in $\{\boldsymbol{f}^l_i\}$ \cite{andriulli_well-conditioned_2013}, with $i=1,...,\NRWGL$ and $j=1,...,\NRWGS$.

The scalar space is separated into the one-dimensional space spanned by the even function $f^e_1 = \text{S}_0^{-1}\left[1\right]$ (where $1$ is the constant function on $\Gamma$ taking on the value $1$ everywhere) and the orthogonal space with respect to the $H^{-\frac{1}{2}}\left(\Gamma\right)$ inner product, spanned by the odd functions $\{f^o_i\}$, $i=1,...,\NPCO$. 
The even function $f^e_1$ is found by solving $\text{Z}^\text{e} \text{f}^e_1 = \boldsymbol{1}$, with $\boldsymbol{1}_i \vcentcolon= \left<f_i,1\right>$ and $\text{Z}^\text{e}_{i,j} \vcentcolon=\left<f_i,S_0\left[f_j\right]\right>_\Gamma$. The discrete coefficients, in the PC space, of $f^e_1$, are denoted as $\text{f}^e_{1,i}$ .

\subsection{Projectors}
The quasi-Helmholtz projectors from \cite{andriulli_well-conditioned_2013} are defined as:   
\begin{align}
    \text{P}_\Sigma &\vcentcolon= \Sigma\left(\Sigma^T\Sigma\right)^\dag\Sigma^T
\\\label{eq:startprojector}
     \text{P}_\Lambda &\vcentcolon= \text{I} - \text{P}_\Sigma
\\
    \mathbb{P}_\Lambda &\vcentcolon= \Lambda\left(\Lambda^T\Lambda\right)^\dag\Lambda^T
    \\
    \mathbb{P}_\Sigma &\vcentcolon= \text{I}-\mathbb{P}_\Lambda
    \\
    \text{P}_\text{GL} &\vcentcolon= \mathbb{P}_\text{GL} \vcentcolon=\mathbb{P}_\Sigma - \text{P}_\Sigma,
\end{align}
where $\Sigma$ is the oriented incidence matrix of the faces and edges, and $\Lambda$ is the oriented incidence matrix of the vertices and the edges, both defined in \cite{andriulli_well-conditioned_2013}.\\ 
The projection on $f^e_1$ is obtained by applying the $H^{-\frac{1}{2}}\left(\Gamma\right)$ inner product $\left(a,b\right)_{H^{-\frac{1}{2}}\left(\Gamma\right)} \vcentcolon= \left<a,S_0\left[b\right] \right>_\Gamma\quad a,b\in H^{-\frac{1}{2}}\left(\Gamma\right)$ as,

\begin{equation}
    \text{P}_{e;i,j} \vcentcolon= \frac{\text{f}^e_{1,i}\left(f^e_1,f_j\right)_{H^{-\frac{1}{2}}\left(\Gamma\right)}}{\left(f^e_1,f^e_1\right)_{H^{-\frac{1}{2}}\left(\Gamma\right)}}
    =\frac{\text{f}^e_{1,i}\left<1,f_j\right>_\Gamma}{\left<f^e_1,1\right>_\Gamma}.
\end{equation}
The projection matrix on the odd functions is defined as
\begin{align}
    \label{eq:mzproj}
        \text{P}_o &\vcentcolon= \text{I}-\text{P}_e.
\end{align}
\subsection{Low-Frequency Scaling}\label{sec:low_freq_scaling}
The matrices from equations \eqref{eq:discvpie-v} and \eqref{eq:disc-vpie-c} are written in terms of the previously defined loop (subscript l) and star (subscript s) subspace for the vectorial degrees of freedom, and even (subscript e) and odd (subscript o) space for the scalar degrees of freedom. The test-space is decomposed similarly.
This yields
\begin{equation}\label{eq:decvecv}\hspace{-10pt}
\setlength{\arraycolsep}{2pt}
    \begin{bmatrix}
    \text{Z}^{11}_{\text{l},\text{l}} & \text{Z}^{11}_{\text{l},\text{s}} & 0_{\NRWGL\times1} & 0_{\NRWGL\times\NPCO} & 0_{\NRWGL\times 1}\\
    \text{Z}^{11}_{\text{s},\text{l}} & \text{Z}^{11}_{\text{s},\text{s}} & \text{Z}^{12}_{\text{s},\text{e}} & \text{Z}^{12}_{\text{s},\text{o}} & Z^{13}_{\text{s},1}\\
    0_{1\times\NRWGL} & \text{Z}^{21}_{\text{e},\text{s}} & \text{Z}^{22}_{\text{e},\text{e}} & \text{Z}^{22}_{\text{e},\text{o}} & \text{Z}^{23}_{\text{e},1}\\
    0_{\NPCO\times\NRWGL} & \text{Z}^{21}_{\text{o},\text{s}} &\text{Z}^{22}_{\text{o},\text{e}} &\text{Z}^{22}_{\text{o},\text{o}} & \text{Z}^{23}_{\text{o},1}\\
    0_{1\times\NRWGL}&0_{1\times\NRWGS}&\text{Z}^{32}_{1,\text{e}}&\text{Z}^{32}_{1,\text{o}}&0_{1\times 1}
    \end{bmatrix}
    \begin{bmatrix}
    \text{v}_\text{l}\\\text{v}_\text{s}\\\text{w}_\text{e}\\\text{w}_\text{o} \\\text{V}_{\boldsymbol{\text{A}}}
    \end{bmatrix}
    =\begin{bmatrix}
        \text{y}_\text{l}\\\text{y}_\text{s}\\\text{z}_\text{e}\\\text{z}_\text{o} \\0
    \end{bmatrix}
\end{equation}
for the VPIE-V method and
\begin{equation}\label{eq:decvecc}
\setlength{\arraycolsep}{2pt}
    \begin{bmatrix}
    \text{Z}^{11}_{\text{l},\text{l}} & \text{Z}^{11}_{\text{l},\text{s}} & 0_{\NRWGL\times1} & 0_{\NRWGL\times\NPCO} \\
    \text{Z}^{11}_{\text{s},\text{l}} & \text{Z}^{11}_{\text{s},\text{s}} & \text{Z}^{12}_{\text{s},\text{e}} & \text{Z}^{12}_{\text{s},\text{o}} \\
    0_{1\times\NRWGL} & \text{Z}^{21}_{\text{e},\text{s}} & \text{Z}^{22}_{\text{e},\text{e}} & \text{Z}^{22}_{\text{e},\text{o}} \\
    0_{\NPCO\times\NRWGL} & \text{Z}^{21}_{\text{o},\text{s}} &\text{Z}^{22}_{\text{o},\text{e}} &\text{Z}^{22}_{\text{o},\text{o}}
    \end{bmatrix}
    \begin{bmatrix}
    \text{v}_\text{l}\\\text{v}_\text{s}\\\text{w}_\text{e}\\\text{w}_\text{o} \\
    \end{bmatrix}
    =\begin{bmatrix}
        \text{y}_\text{l}\\\text{y}_\text{s}\\\text{z}_\text{e}\\\text{z}_\text{o} 
    \end{bmatrix}
\end{equation}
for the VPIE-C method, where $\text{v}_\text{l}$, $\text{v}_\text{s}$, $\text{w}_\text{e}$, and $\text{w}_\text{o}$ represent the coefficient vectors in the loop, star, even, and odd subspaces, respectively.\\
Each matrix or vector element in \eqref{eq:decvecv} and \eqref{eq:decvecc} exhibits a specific dependence on the angular frequency $\omega$. At low frequencies, we can approximate their frequency behavior by a Laurent series around $\omega=0$.
The $\left(a,b\right)$ notation describes, for both the real and imaginary part, the lowest-order non-zero term in this Laurent series, i.e. if a matrix O is dependent on $\omega$ as $\left(a,b\right)$, this means that $Re\left(\text{O}\right) \propto \omega^a$ and $Im\left(\text{O}\right) \propto \omega^b$ when $\omega \to 0$.
The $\text{Z}^{12}_{\text{s},\text{e}}$ block can be written as a contribution in $\kappa = 0$ and a higher-order contribution. The explicit expression for the star to RWG map $\Sigma$, and even function to PC map ${\text{Z}^\text{e}}^{-1}$ have been inserted, which gives
\begin{align}\label{eq:Zsoprop}
   \text{Z}^{12}_{\text{s},\text{e}}  &= \Sigma^T \text{Z}^\text{e} {\text{Z}^\text{e}}^{-1} \boldsymbol{1} + \mathcal{O}\left(2,3\right) = \mathcal{O}\left(2,3\right),
\end{align}
where we used $\Sigma^T \boldsymbol{1} = 0$. This relation is used to derive the scaling order of \eqref{eq:decvecv} and \eqref{eq:decvecc}.

The following scaling orders for the VPIE-V and VPIE-C methods are found, respectively:
\begin{equation} \hspace*{-11pt}   
\setlength{\arraycolsep}{2pt}
    \begin{bmatrix}
\left(0,1\right) & \left(0,1\right) &0&0&0\\
\left(0,1\right)&\left(0,1\right)&\left(2,3\right)&\left(0,1\right)&\left(0,1\right)\\
0&\left(2,3\right)&\left(2,3\right)&\left(2,3\right)&\left(0,1\right)\\
0&\left(0,1\right)&\left(2,3\right)&\left(2,3\right)&\left(0,1\right)\\ 0&0&\left(0,/\right)&\left(0,/\right)&0
    \end{bmatrix}\hspace*{-5pt} \begin{bmatrix}
    \mathcal{O}\left(\text{v}_\text{l}\right)\\\mathcal{O}\left(\text{v}_\text{s}\right)\\\mathcal{O}\left(\text{w}_\text{e}\right)\\\mathcal{O}\left(\text{w}_\text{o}\right)\\\mathcal{O}\left(\text{V}_{\boldsymbol{A}}\right)
    \end{bmatrix} = \begin{bmatrix}
        \left(0,1\right)\\\left(0,1\right)\\\left(2,1\right)\\\left(2,1\right)\\0
    \end{bmatrix}
    \label{eq:scal_VS_VPIE_V}
\end{equation}
\begin{equation} \hspace*{-8pt}
\setlength{\arraycolsep}{2pt}
    \begin{bmatrix}
\left(0,1\right) & \left(0,1\right) &0&0\\
\left(0,1\right)&\left(0,1\right)&\left(2,3\right)&\left(0,1\right)\\
0&\left(2,3\right)&\left(2,3\right)&\left(2,3\right)\\
0&\left(0,1\right)&\left(2,3\right)&\left(2,3\right)
    \end{bmatrix}\hspace*{-5pt} \begin{bmatrix}
    \mathcal{O}\left(\text{v}_\text{l}\right)\\\mathcal{O}\left(\text{v}_\text{s}\right)\\\mathcal{O}\left(\text{w}_\text{e}\right)\\\mathcal{O}\left(\text{w}_\text{o}\right)
    \end{bmatrix} = \begin{bmatrix}
        \left(0,1\right)\\\left(0,1\right)\\\left(2,1\right)\\\left(2,1\right)
    \end{bmatrix}.
    \label{eq:scal_VS_VPIE_C}
\end{equation}
The scaling for the degrees of freedom can be obtained immediately by observing the matrix and the right hand side in \eqref{eq:scal_VS_VPIE_V} and \eqref{eq:scal_VS_VPIE_C}. This yields 
\begin{subequations}
\label{eq:scalingdof-V}
\begin{align}\mathcal{O}\left(\text{v}_\text{l}\right) &= \left(0,1\right)\\
\mathcal{O}\left(\text{v}_\text{s}\right) &= \left(2,1\right)\\
\mathcal{O}\left(\text{w}_\text{e}\right) &= \left(0,1\right)\\
\mathcal{O}\left(\text{w}_\text{o}\right) &= \left(0,1\right)\\
\mathcal{O}\left(\text{V}_{\boldsymbol{A}}\right) &= \left(2,1\right)
\end{align}
\end{subequations}for the VPIE-V and 
\begin{subequations}
\label{eq:scalingdof-C}
\begin{align}
\mathcal{O}\left(\text{v}_\text{l}\right) &= \left(0,1\right)\\
\mathcal{O}\left(\text{v}_\text{s}\right) &= \left(2,1\right)\\
\mathcal{O}\left(\text{w}_\text{e}\right) &= \left(0,-1\right) \label{eq:dofscalingwe} \\
\mathcal{O}\left(\text{w}_\text{o}\right) &= \left(0,1\right)
\end{align} 
\end{subequations}for the VPIE-C. The difference in scaling is expected as both methods are subject to a different choice of gauge.
However, errors are made in a numerical context because of the finite precision of integration routines, floating point representations and (iterative) matrix inversion algorithms, because of which those exact scalings are not obtained. Instead, those errors typically follow a different scaling in $\omega$, making them significantly larger compared to the corresponding components of the discrete solution in absence of numerical errors. As a result, these solution components are not computed correctly and any physically relevant quantity in which they dominate can no longer be retrieved. To stabilize the method with respect to those errors, a dedicated procedure is required, as explained in the next section. The influence of the numerical errors on the actual scaling of the degrees of freedom is further discussed in the result section.
\subsection{Stabilization Procedure}
The matrix, the right-hand-side (RHS) and the degrees of freedom (DOF) are each equipped with their own stabilization scheme to obtain a broadband accurate, final system.
The stabilization procedure for those three parts is discussed in further detail, after which they are combined into a single scheme.
\subsubsection{Matrix}
The stabilization of the matrix focuses on setting contributions that are zero explicitly to zero to eliminate some level of quadrature and round-off error. The choice of the scaling is covered by the RHS and DOF rescaling.
The contributions $\text{Z}^{12}_{\text{l},\text{e}}, \text{Z}^{12}_{\text{o}}$, $\text{Z}^{21}_{\text{e},\text{l}}$ and  $\text{Z}^{21}_{\text{o},\text{l}}$ need to be set explicitly to zero. because the finite precision of the quadrature rules induces an error that scales with a lower order in $\omega$, and will become the main contribution at low frequencies. For the VPIE-C method, it was derived in \eqref{eq:dofscalingwe} that the $\text{w}_\text{e}$ component scales as $\left(0,-1\right)$ at low frequencies. Therefore, we have to set the lowest order terms of $\text{Z}^{12}_{\text{s},\text{e}}$ and $\text{Z}^{21}_{\text{e},\text{s}}$ explicitly to zero as we cannot allow a small but constant round-off error (with respect to $\omega$). This is not necessary in the VPIE-V method, because the components $\text{w}_\text{e}$ and $\text{w}_\text{o}$ exhibit the same low-frequency behavior. The error induced by not setting the contribution explicitly to zero is in this case not magnified by the $\text{w}_\text{e}$ component.

The higher-order Green function, defined as ${G^{\left(2,3\right)}\left(\boldsymbol{r},\boldsymbol{r'}\right) \vcentcolon= G-G_0+\frac{\jmath\kappa}{4\pi}} = \frac{e^{-\jmath\kappa|\boldsymbol{r}-\boldsymbol{r'}|}-1+\jmath\kappa|\boldsymbol{r}-\boldsymbol{r'}|}{4\pi |\boldsymbol{r}-\boldsymbol{r'}|}$, is introduced to compute operators in which the lowest order term is set to zero. Matrices computed with this Green function are denoted with a tilde, we define
\begin{align}
    \widetilde{\text{Z}}^{12}_{i,j} &\vcentcolon= -\left<\nabla\cdot\boldsymbol{f}_i, S^{(2,3)}_\kappa\left[f_j\right]\right>_\Gamma \\
    \widetilde{\text{Z}}^{21}_{i,j} &\vcentcolon= \widetilde{\text{Z}}^{12}_{j,i}. 
\end{align}
\subsubsection{Degrees of Freedom}
In the degrees of freedom, one should be careful that loop, star, even, or odd contributions with a different scaling in $\omega$ are not stored in the same floating point variable because of round-off and numerical errors. If the star contribution, scaling with $\kappa$, becomes smaller than the loop-dominated numerical error, the error may be misidentified as a star contribution. Contributions are therefore rescaled to remove potential errors introduced by this phenomenon. Setting $\text{v}_\text{s} = \jmath\kappa \text{v}'_\text{s}$ and solving for $\text{v}_\text{s}'$, removes the issue of cancellation \cite{andriulli_well-conditioned_2013}. The scalar unknown is treated in the same manner in the case of the VPIE-C method, yielding $\text{w}_\text{e} = \frac{1}{\jmath\kappa} \text{w}_\text{e}'$. 
\subsubsection{Right-Hand-Side}
The stabilization of the right-hand side is performed for two reasons: First, the same floating point cancellation occurs that affects the left-hand-side. However, this is not an issue for plane waves. For more general excitations we refer to similar techniques as described in \cite{Excitation-Aware-Bernd}. Second, rescaling the right-hand side corresponds to rescaling the equations and ensuring that they still allow for a unique solution at zero frequency, eliminating possible nullspaces. 

\subsubsection{Stabilized Schemes}
The previously discussed stabilizations are applied by exploiting the projectors defined in \eqref{eq:startprojector}-\eqref{eq:mzproj}, which give access to the loop, star, even, and odd components. The final stabilized VPIE-C method is called the low-frequency VPIE-C (LF-VPIE-C) and is given by
\begin{equation}
\hspace*{-11pt}   
\setlength{\arraycolsep}{0pt}
     \begin{bmatrix}
        \text{Z}^{11}\left(\text{P}_\Lambda+\jmath\kappa\text{P}_\Sigma\right) & \text{P}_\Sigma \left(\frac{1}{\jmath\kappa}\widetilde{\text{Z}}^{12}\text{P}_\text{e}+\text{Z}^{12}\text{P}_\text{o}\right)\\
        \left(\text{P}^T_\text{e} \widetilde{\text{Z}}^{21} + \text{P}^T_\text{o}\text{Z}^{21}\right)\text{P}_\Sigma & \frac{1}{\jmath\kappa}\text{Z}^{22} \left(\frac{1}{\jmath\kappa}\text{P}_\text{e}+\text{P}_\text{o}\right)
       
    \end{bmatrix}\begin{bmatrix}\text{v}'\\\text{w}'\\\end{bmatrix} = 
    \begin{bmatrix}
        \text{y} \\ \frac{1}{\jmath\kappa}\text{z}
    \end{bmatrix}
\end{equation}
with $\text{v} = \left(\text{P}_\Lambda + \jmath\kappa\text{P}_\Sigma\right)\text{v}'$  and $\text{w} = \left(\frac{1}{\jmath\kappa}\text{P}_\text{e}+\text{P}_\text{o}\right)\text{w}'$.
The final stabilized VPIE-V method is called the LF-VPIE-V and is given by
\begin{equation}
    \begin{bmatrix}
        \text{Z}^{11}\left(\text{P}_\Lambda+\jmath\kappa\text{P}_\Sigma\right) & \text{P}_\Sigma\text{Z}^{12} & \jmath\kappa\text{Z}^{13}\\
        \text{Z}^{21}\text{P}_\Sigma & \frac{1}{\jmath\kappa}\text{Z}^{22} & \text{Z}^{23}\\
        0 & \text{Z}^{32} & 0
    \end{bmatrix} \begin{bmatrix} \text{v}'\\\text{w}\\\text{V}_{\boldsymbol{A}}'
    \end{bmatrix} = \begin{bmatrix}
        \text{y}\\\frac{1}{\jmath\kappa}\text{z}\\0
    \end{bmatrix}
\end{equation}
with $\text{v} = \left(\text{P}_\Lambda + \jmath\kappa\text{P}_\Sigma\right)\text{v}'$ and $\text{V}_{\boldsymbol{A}} = \jmath\kappa\text{V}'_{\boldsymbol{A}}$.

\section{Calderón Preconditioning}
\subsection{Preconditioner}
A Calderón multiplicative preconditioner is designed to overcome the dense-mesh breakdown. The cue for this left preconditioner is taken from \cite{hawkins_analytic_2023} and \cite{GlobalMultiCaldPrec}, inspired by the Calderón identities of the vector potential. In particular in \cite{HIPTMAIR2006699} it is shown that, given the mapping properties of the continuous and discrete operators, the spectral condition number is bounded independent of the mesh density $h$. A large amount of numerical evidence lead to the conclusion that this is a sufficient condition for fast convergence in an iterative Krylov solver. We will define the preconditioner for the (LF)-VPIE-C method and in the end of the section extend it to the (LF)-VPIE-V method. 
The Calderón preconditioner is given by the discretized version of the $SL^{\kappa ND}$ operator from \cite{GlobalMultiCaldPrec}.

The discrete spaces used to construct the preconditioning matrix are supported by the dual mesh and are given by the Buffa-Christiansen (BC) basis functions, here denoted as $\boldsymbol{b}_i$ and the dual first-order Lagrange (DL1) elements from \cite{buffa_dual_2005}, denoted as $b_i$. Those dual spaces are used such that the dimension of the preconditioner matches the dimension of the primal system. Besides in \cite{HIPTMAIR2006699} it is shown that there should exist a stable duality pairing between the discrete spaces used for the discretization of the primal system and the preconditioner. The preconditioning matrix $\text{Q}$ is given by \cite{GlobalMultiCaldPrec}
\begin{equation}
    \text{Q} = \begin{bmatrix} \text{Q}^{11}_0 + \text{Q}^{11}_2& \text{Q}^{12}_0\\
    \text{Q}^{21}_0 & \text{Q}^{22}_0 \\
\end{bmatrix}
\end{equation}
with
\begin{align}
\text{Q}^{11}_{0,ij} &\vcentcolon= \left<\nabla\cdot\boldsymbol{b}_i,  S_\kappa\left[ \nabla\cdot\boldsymbol{b}_j\right]\right>_\Gamma\\
    \text{Q}^{11}_{2,ij} &\vcentcolon= -\kappa^2\left<\boldsymbol{b}_i,  S_\kappa\left[ \boldsymbol{b}_j\right]\right>_\Gamma\\
    \text{Q}^{21}_{0,ij} &\vcentcolon= -\left<\overline{\operatorname{curl}_\Gamma}\left(b_i\right), S_\kappa\left[\boldsymbol{b}_j\right]\right>_\Gamma\\
    \text{Q}^{12}_{0,ij} &\vcentcolon= -\text{Q}^{21}_{0,ji}\\
    \text{Q}^{22}_{0,ij} &\vcentcolon= \left<b_i, \tau_N\circ S_\kappa\left[ \boldsymbol{n}'b_j\right]\right>_\Gamma.
\end{align}
\subsection{Low-Frequency Behavior}
The low-frequency behavior of the preconditioner is studied in the same way as for the primal system. The BC space is decomposed into a local loop space spanned by $\left\{\boldsymbol{b}_i^{l'}\right\}_i$, a local star space spanned by $\left\{\boldsymbol{b}_i^{s}\right\}_i$ and the space spanned by the global loops, which has a dimension equal to twice the genus of the object, spanned by $\left\{\boldsymbol{b}_i^{gl}\right\}_i$. The scalar dual Lagrange space is decomposed into the one-dimensional space spanned by the constant function on $\Gamma$, $b_1^c$, and the orthogonal complement with respect to the $L_2\left(\Gamma\right)$ inner product, spanned by the functions that have mean zero, i.e. $\left\{b_i|\int_\Gamma b_i \quad dA= 0\right\}$. Let $b^{mz}$ be a basis for this space, the decomposed matrix is then given by
\begin{equation}\label{eq:precdec}
\setlength{\arraycolsep}{2pt}
    \tilde{\text{Q}} = \left[\begin{array}{ccc|cc}
        \text{Q}_{2,l',l'}^{11} & \text{Q}_{2,l',gl}^{11} & \text{Q}_{2,l',s}^{11} & 0 & \text{Q}_{0,l',mz}^{12} \\
        \text{Q}_{2,gl,l'}^{11} & \text{Q}_{2,gl,gl}^{11} & \text{Q}_{2,gl,s}^{11} & 0 & \text{Q}_{0,gl,mz}^{12} \\
        \text{Q}_{2,s,l'}^{11} & \text{Q}_{2,s,gl}^{11} & \text{Q}_{0,s,s}^{11}+\text{Q}_{2,s,s}^{11}& 0 &\text{Q}_{0,s,mz}^{12} \\\hline
       0&0&0 & \text{Q}_{0,c,c}^{22} & \text{Q}_{0,c,mz}^{22} \\
        \text{Q}_{0,mz,l'}^{21} & \text{Q}_{0,mz,gl}^{21} & \text{Q}_{0,mz,s}^{21} & \text{Q}_{0,mz,c}^{22} & \text{Q}_{0,mz,mz}^{22} 
    \end{array}\right].
\end{equation}
The low-frequency scaling of each block is given by
\begin{equation}\hspace*{-11pt}   
     \mathcal{O}\left(\tilde{\text{Q}}\right)= \left[\begin{array}{ccc|cc}
        \left(2,3\right) & \left(2,3\right) &\left(2,3\right) & 0 & \left(0,1\right) \\
        \left(2,3\right) &\left(2,3\right) &\left(2,3\right) & 0 & \left(0,1\right)  \\
        \left(2,3\right) &\left(2,3\right) &\left(0,1\right) & 0 & \left(0,1\right) \\\hline
        0 & 0 & 0 & \left(0,1\right) & \left(0,1\right) \\
        \left(0,1\right) &\left(0,1\right)  &\left(0,1\right) &\left(0,1\right) &\left(0,1\right)
    \end{array}\right].
\end{equation}
When $\omega = 0$, we see that the nullspace of the $\text{Q}^{11}$ block has a dimension equal to the total number of local and global loops. The $\text{Q}^{12}$ block has a rank equal to the number of DL1 elements minus one (for the constant element), which equals the total number of local loops. The concatenated $\text{Q}^{11}$ with $\text{Q}^{21}$ block, given by the $\text{Q}^{:1}$ block, will therefore have a nullspace with dimension equal to at least the number of global loops, which is twice the genus of the object. This nullspace can be removed by adding a judiciously chosen low-rank matrix, following the cue from a similar technique used in \cite{low_rank_prec}. This is allowed because adding a finite-rank contribution to the preconditioner will not influence the final solution.  \\
Two points remain to be proven, first that this nullspace is equal to the number of global loops, and second, that the constructed low-rank matrix "covers" the nullspace of the preconditioner.
From the $\text{Q}^{11}$ block we already know that the nullspace of the $\text{Q}^{:1}$ block is contained within the direct sum of the local and global loop space.
The nullspace can thus be found by solving 
\begin{equation}
    \text{Q}^{21}_{0,mz,l'} \text{u}_{l'} + \text{Q}^{21}_{0,mz,gl}\text{u}_{gl} = 0.
\end{equation}
If an element in the nullspace is $l_2$ orthogonal to the global loop elements, i.e. purely in the local loop space, we would have 
\begin{equation}
    \text{Q}^{21}_{0,mz,l'} \text{u}_{l'}  = 0.
\end{equation}
which indicates that the single-layer operator assembled in the local loop space is not full rank and thus not positive definite. This contradicts the fact that there exists an element in this nullspace that is $l_2$ orthogonal to the global loop space. This also indicates that the dimension of the nullspace of $\text{Q}^{:1}$ is equal to the number of global loops.
We can thus remove the nullspace of the preconditioner by adding the following low-rank matrix containing the $l_2$ projectors on the global loop space
\begin{equation}
    \text{Q}_\text{LR} = \begin{bmatrix}\text{P}_\text{GL}\text{G}_\text{BC}\text{P}_\text{GL} & 0_{\NRWG \times \NPC}\\
    0_{\NPC\times\NRWG} & 0_{\NPC\times\NPC}
    \end{bmatrix},
\end{equation}
in which the Gramm matrix on the dual BC space is defined as 
\begin{equation}
    \text{G}_{\text{BC},ij} = \left<\boldsymbol{b}_i,\boldsymbol{n}\times\boldsymbol{b}_j\right>_\Gamma .
\end{equation}
The complete left-preconditioner for the (LF)-VPIE-C methods is given by
\begin{equation}
    \text{Q}_\text{C} \vcentcolon= \begin{bmatrix}-\text{G}^{-1}_\text{BC,RT} & 0\\ 0 & \text{G}^{-1}_\text{DL1,PC}\end{bmatrix}\left(\text{Q}+\text{Q}_{\text{LR}}\right)\begin{bmatrix}\text{G}^{-1}_\text{BC,RT} & 0\\ 0 & \text{G}^{-1}_\text{DL1,PC}\end{bmatrix}^T,
\end{equation}
with
\begin{align}
    \text{G}_\text{BC,RT,ij} &\vcentcolon= \left<\boldsymbol{b}_i,
    \boldsymbol{n}\times\boldsymbol{f}_j\right>\\
    \text{G}_\text{DL1,PC,ij} &\vcentcolon= \left<b_i,f_j\right>.
\end{align}
The left-preconditioner for the (LF)-VPIE-V method is given by
\begin{equation}
    \text{Q}_\text{V} = \begin{bmatrix}
        \text{Q}_C & 0\\0& 1
    \end{bmatrix}.
\end{equation}
\section{Results}\label{sec:results}
\begin{figure}
    \centering
    \includegraphics[width=\linewidth]{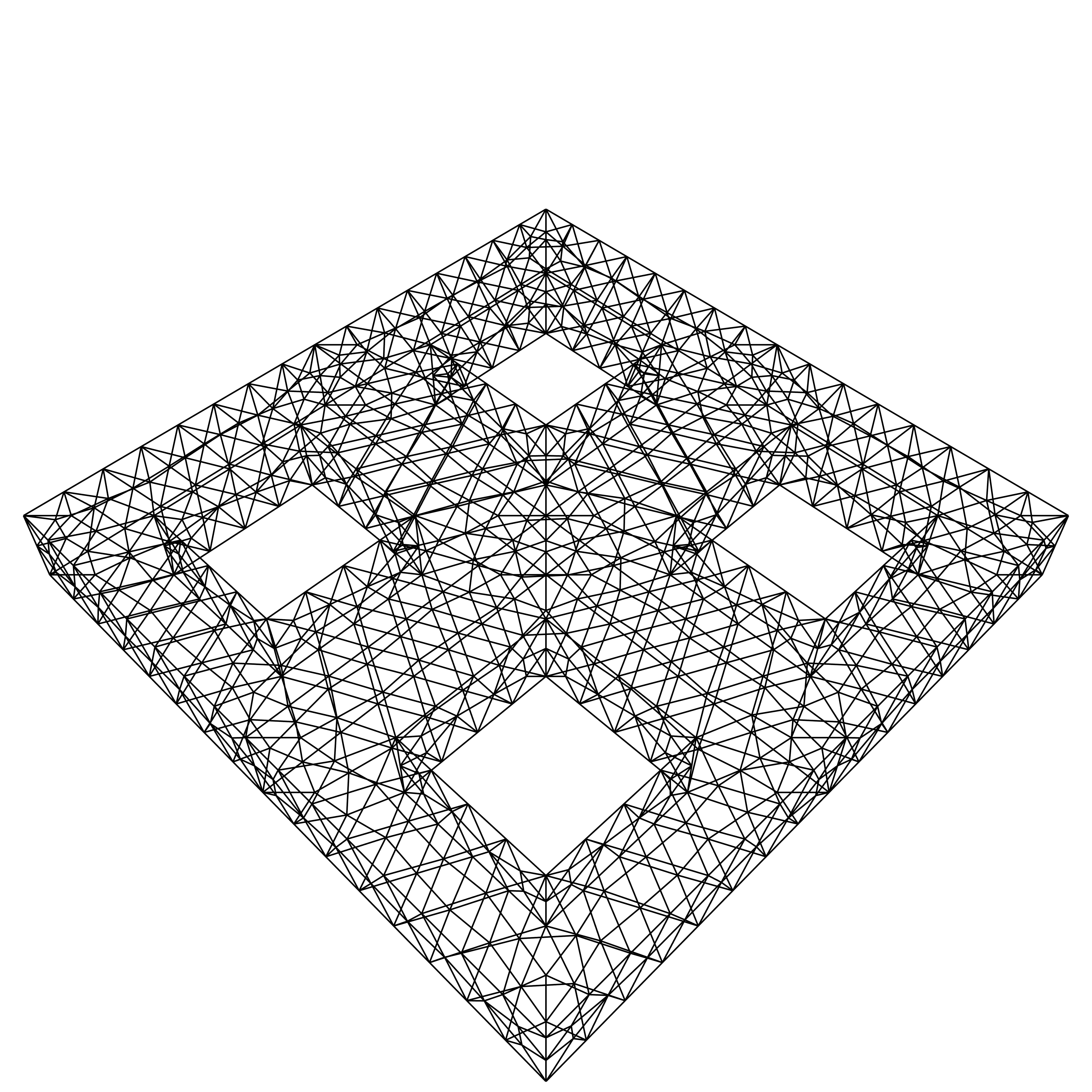}
    \caption{Triangulated mesh of the 2x2 torus with side 8~m. The holes have side 2~m. The height is 1~m.}
    \label{fig:holes_mesh}
\end{figure}

The plane-wave scattering on a 2×2 torus (Fig. \ref{fig:holes_mesh}) is tested using the four (LF)-VPIE-(C/V) methods. More specifically, we examine the low-frequency behavior of the degrees of freedom as a function of this frequency. From this analysis, we determine the physical quantities that are correctly estimated by the non-stabilized methods and the ones that require the stabilized methods.\\
Afterward, it is shown that the number of iterations in the iterative Krylov solver of the Calderón-preconditioned systems does not suffer from dense-mesh breakdown or, as is well known in the literature, suffer from low-frequency breakdown \cite{vico_decoupled_2016}.
\subsection{Scaling of Degrees of Freedom}
We consider only the leading terms of the scaling of the degrees of freedom and physical quantities and, thereby, no distinction is made between the scaling of the real and imaginary part as was done in the theory. This is done for two reasons: first, because a complex phase shift on the incident wave would remove the higher-order scaling information, for example multiplying both sides of the system with $e^{i\phi}$ mixes up the real and imaginary parts of the degrees of freedom leaving only information about the lowest-order scaling.
Second, the higher-order scaling is not detected by the convergence criteria of the iterative Krylov solver.
However, taking only the lowest-order contribution into account does not introduce any issues in the accurate computation of the near and far fields, because it is the decomposition into the Helmholtz components that is important rather than the decomposition into real and imaginary parts.

\begin{figure}[htbp]
\centering

\begin{subfigure}{0.48\columnwidth}
\includegraphics[width=\linewidth]{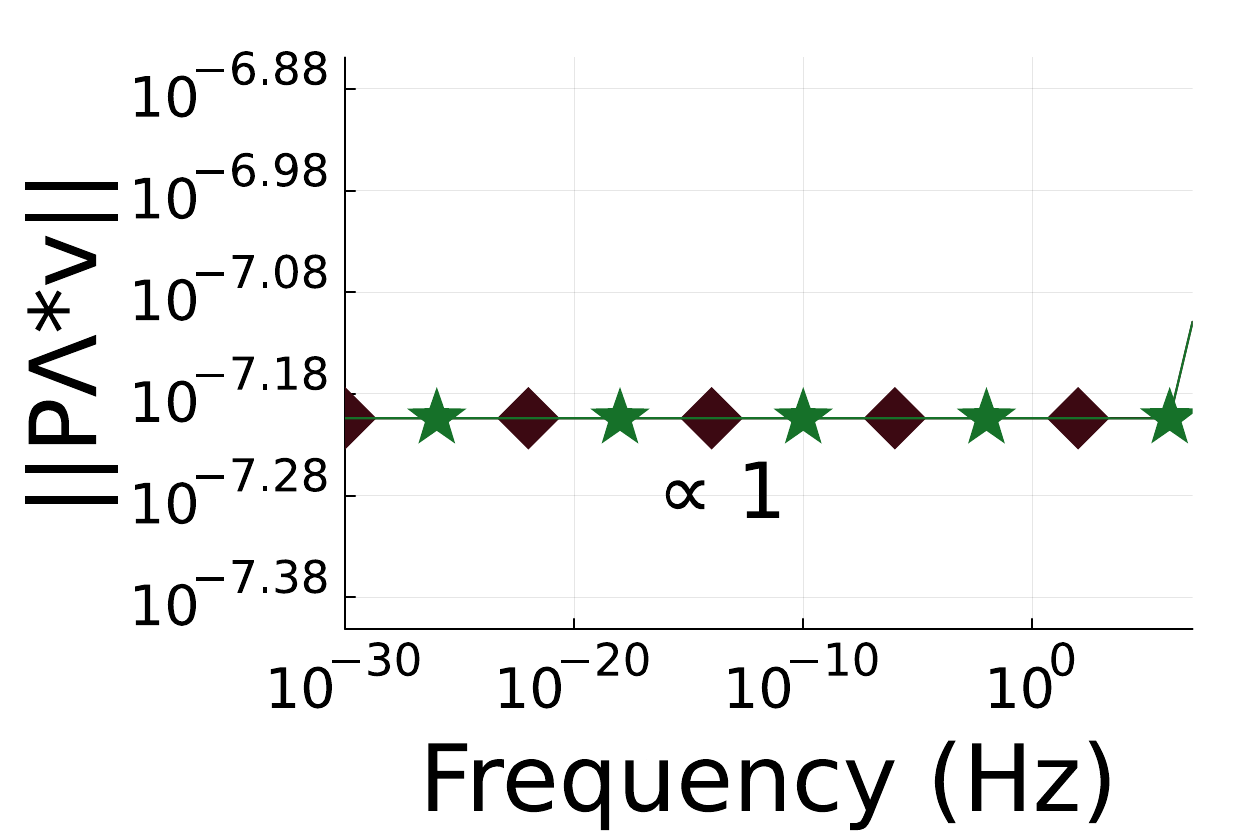}
\caption{$||\text{P}_\Lambda \text{v}||$}
\end{subfigure}
\hfill
\begin{subfigure}{0.48\columnwidth}
\includegraphics[width=\linewidth]{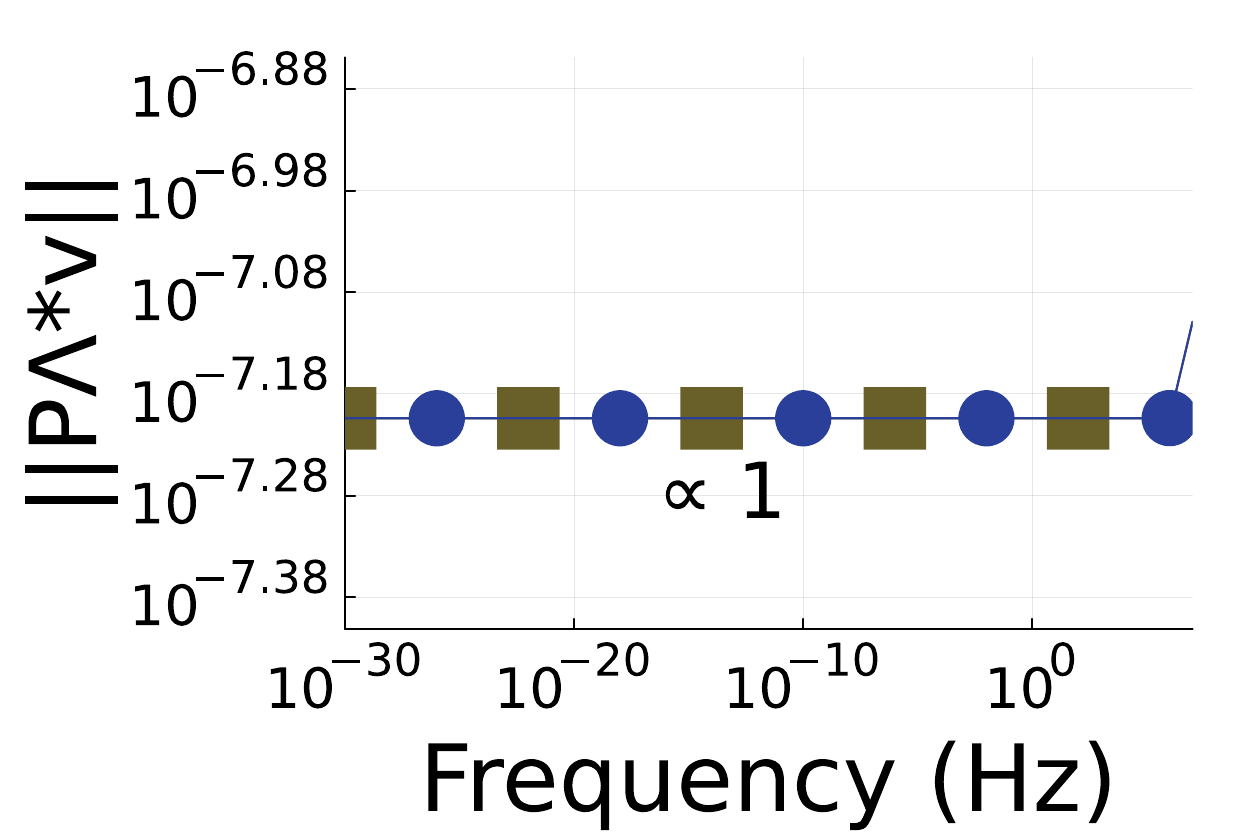}
\caption{$||\text{P}_\Lambda \text{v}||$}
\end{subfigure}

\vspace{0.2cm}

\begin{subfigure}{0.48\columnwidth}
\includegraphics[width=\linewidth]{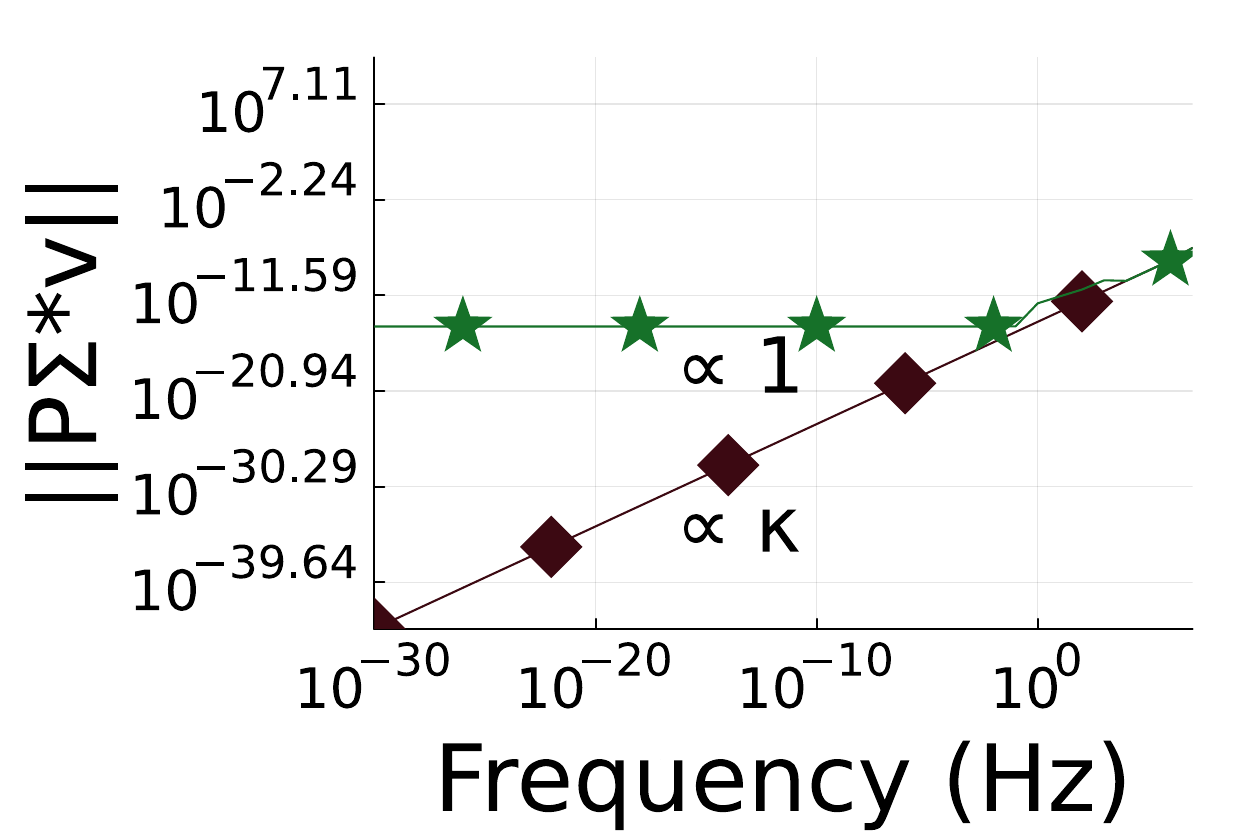}
\caption{$||\text{P}_\Sigma \text{v}||$}
\end{subfigure}
\hfill
\begin{subfigure}{0.48\columnwidth}
\includegraphics[width=\linewidth]{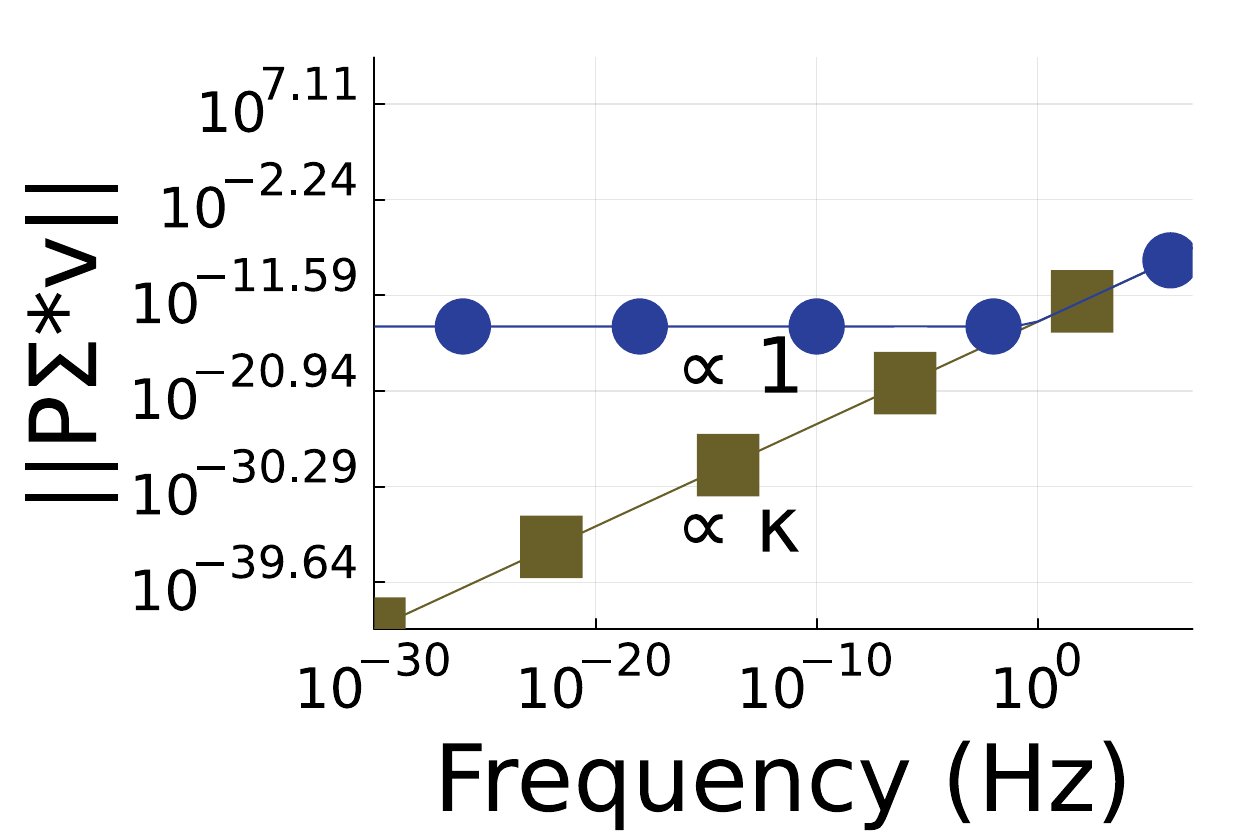}
\caption{$||\text{P}_\Sigma \text{v}||$}
\end{subfigure}

\vspace{0.2cm}

\begin{subfigure}{0.48\columnwidth}
\includegraphics[width=\linewidth]{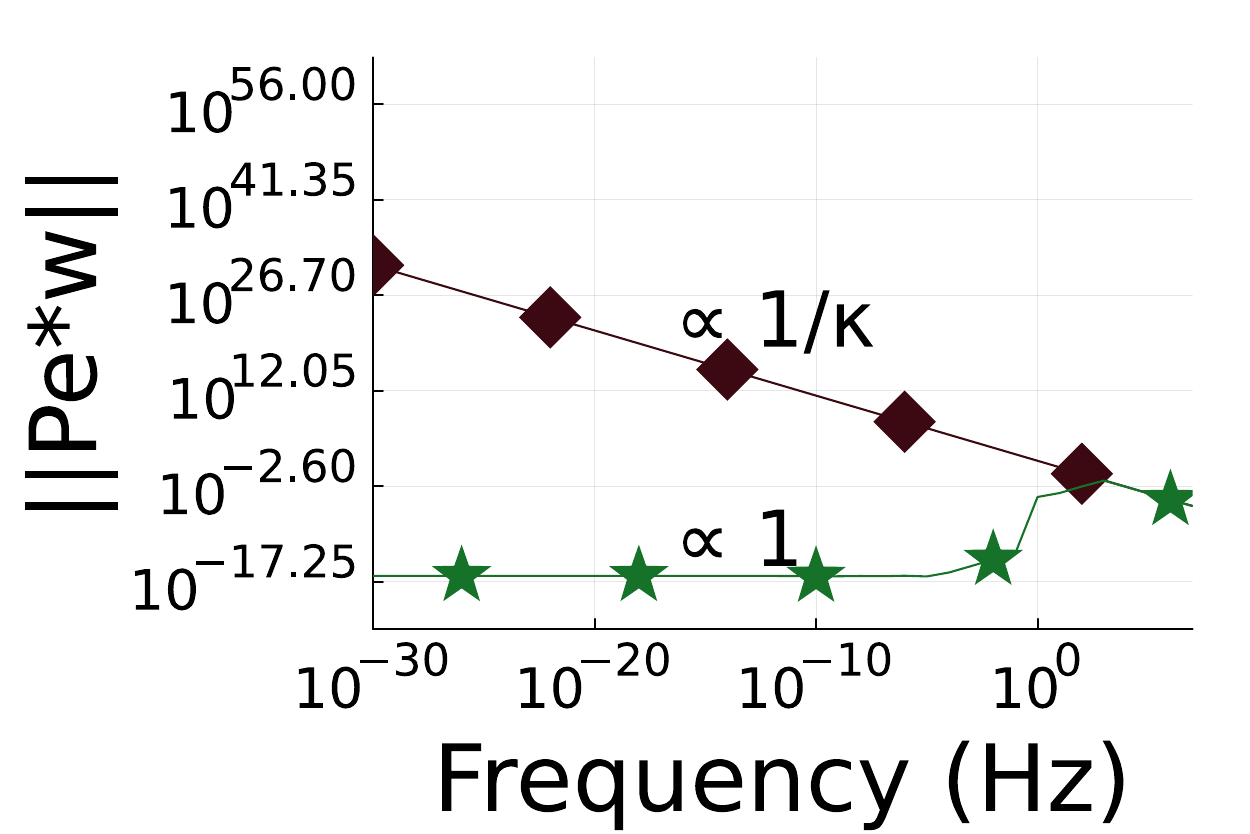}
\caption{ $||\text{P}_e \text{w}||$ }
\end{subfigure}
\hfill
\begin{subfigure}{0.48\columnwidth}
\includegraphics[width=\linewidth]{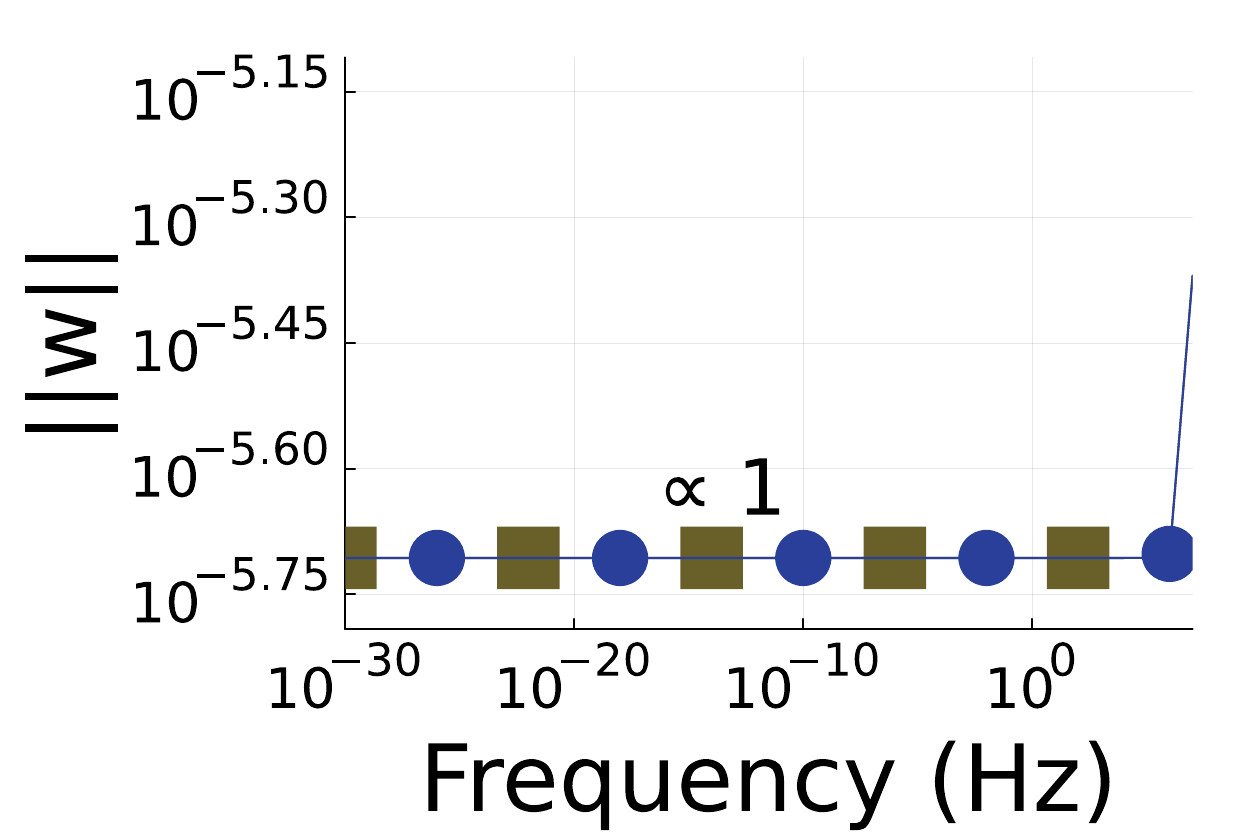}
\caption{ $||\text{w}||$}
\end{subfigure}

\vspace{0.2cm}

\begin{subfigure}{0.48\columnwidth}
\includegraphics[width=\linewidth]{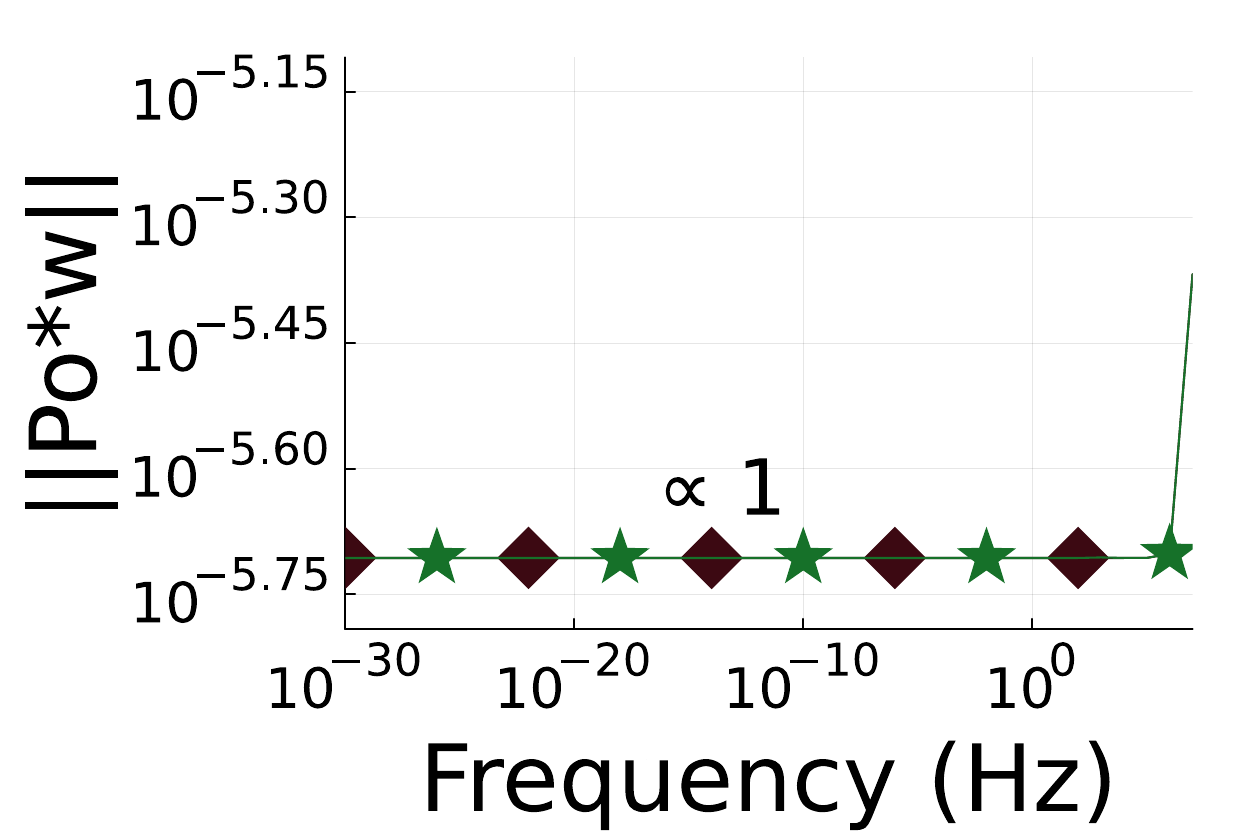}
\caption{ $||\text{P}_o \text{w}||$ }
\end{subfigure}
\hfill
\begin{subfigure}{0.48\columnwidth}
\includegraphics[width=\linewidth]{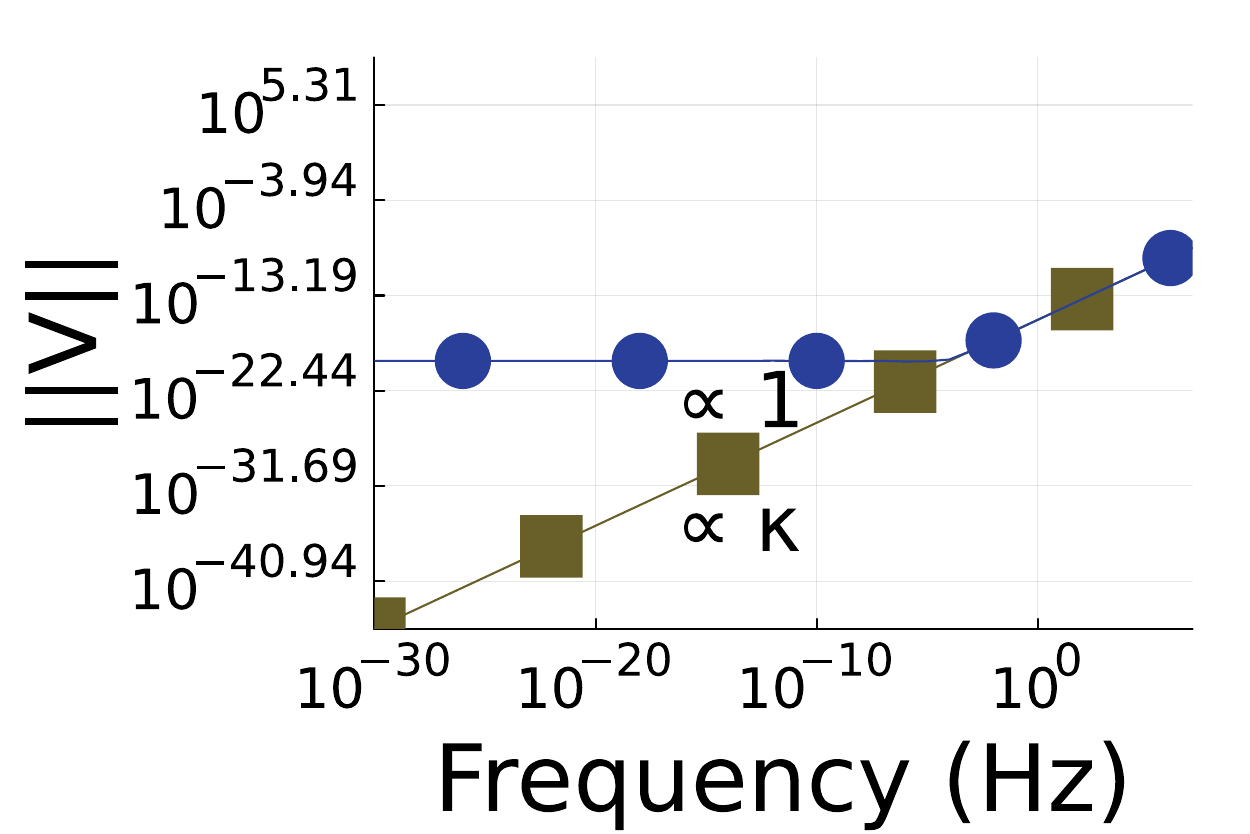}
\caption{ $||\text{V}_{\boldsymbol{A}}||$ }
\end{subfigure}

\vspace{0.3cm}

\begin{subfigure}{\columnwidth}
\centering
\includegraphics[width=0.7\linewidth]{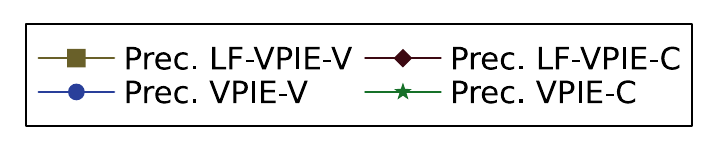}
\end{subfigure}

\caption{Scaling of the norm of the degrees of freedom for the (LF)-VPIE-(C/V) methods. The scaling for the LF methods is as expected from \eqref{eq:scalingdof-V} and \eqref{eq:scalingdof-C}.}
\label{fig:scaling_dof}
\end{figure}

The norm of the different degrees of freedom for the (LF)-VPIE-C methods is shown in Figs \ref{fig:scaling_dof} (left column). The scaling of the degrees of freedom of the stabilized method is as expected from theory. The scaling results of these figures are summarized in Table \ref{tab:scaling_DOF_C} and are used in the next section in the computation of the physical quantities.

Similarly, the results of the DOF-scaling of the (LF)-VPIE-V methods are shown in Figs. \ref{fig:scaling_dof} (right column) and summarized in Table \ref{tab:scaling_DOF_V}.

\begin{table}
    \centering
    \begin{tabular}{c|c|c|c|c}
    Method & $||\text{P}_\Lambda \text{v}||$&$||\text{P}_\Sigma\text{v}||$ & $||\text{P}_\text{e}\text{w}||$&$||\text{P}_\text{o}\text{w}||$ \\\hline
     VPIE-C    &  $\propto 1$&$\propto 1$&$\propto 1$&$\propto 1$\\
     LF-VPIE-C    &$\propto 1$&$\propto \kappa$&$\propto \frac{1}{\kappa}$&$\propto 1$ \\
    \end{tabular}
    \caption{The frequency scaling of the different degrees of freedom at low-frequency for the (LF)-VPIE-C methods.}
    \label{tab:scaling_DOF_C}
\end{table}
\begin{table}
    \centering
    \begin{tabular}{c|c|c|c|c}
    Method & $||\text{P}_\Lambda \text{v}||$&$||\text{P}_\Sigma\text{v}||$ & $||\text{w}||$ &$||\text{V}_{\boldsymbol{A}}||$\\\hline
     VPIE-V & $\propto 1$&$\propto 1$&$\propto 1$&$\propto 1$\\
     LF-VPIE-V & $\propto 1$&$\propto \kappa$&$\propto 1$&$\propto \kappa$\\
    \end{tabular}
    \caption{The frequency scaling of the different degrees of freedom at low-frequency for the (LF)-VPIE-V methods.}
    \label{tab:scaling_DOF_V}
\end{table}
\subsection{Scaling Physical Quantities}
\subsubsection{Far-Field}
The scattered vector-potential is analyzed in the far field regime, where $\kappa r \to \infty$. The frequency is chosen to be small but fixed, ensuring that the far field limit can still be computed. The following far field approximations are applied to the Green function in \eqref{eq:greenfunction}
\begin{equation}
    G\left(\boldsymbol{r},\boldsymbol{r'}\right) \vcentcolon= \frac{e^{-\jmath\kappa|\boldsymbol{r}-\boldsymbol{r'}|}}{4\pi |\boldsymbol{r}-\boldsymbol{r'}|} \stackrel{r >> r' }{\approx} \frac{e^{-\jmath\kappa r}}{4\pi r}e^{\jmath\kappa\boldsymbol{1}_r\cdot\boldsymbol{r'}}
\end{equation}
\begin{equation}
    \nabla G\left(\boldsymbol{r},\boldsymbol{r'}\right) \stackrel{r >> r' }{\approx} -\jmath\kappa\frac{e^{-\jmath\kappa r}}{4\pi r}e^{\jmath\kappa\boldsymbol{1}_r\cdot\boldsymbol{r'}}\boldsymbol{1}_r = -\jmath\kappa G\left(\boldsymbol{r},\boldsymbol{r'}\right) \boldsymbol{1}_r ,
\end{equation}
in which $\boldsymbol{1}_r \vcentcolon= \frac{\boldsymbol{r}}{r}$, yielding
\begin{equation}\label{eq:Aff}
        \boldsymbol{A}^{FF} \vcentcolon= \frac{e^{-\jmath\kappa r}}{4\pi r}\left(\mathcal{S}^{FF}_{\kappa}\left[\boldsymbol{v}\right] +\jmath\kappa\mathcal{S}^{FF}_{\kappa}\left[w\right]\boldsymbol{1_r} - \mathcal{S}^{FF}_{\kappa}\left[\boldsymbol{n}V_{\boldsymbol{A}}\right]\right),
\end{equation}
with the far field single-layer potential defined as
\begin{equation}
    \mathcal{S}^{FF}_\kappa\left[b\right]\left(\boldsymbol{r}\right) \vcentcolon= \int_\Gamma e^{\jmath\kappa\boldsymbol{1_r}\cdot\boldsymbol{r'}} b\left(\boldsymbol{r'}\right)\quad dA,
\end{equation}
where $b$ can be either a scalar- or vector-valued function defined on $\Gamma$. The lowest-order term of this far-field single-layer operator is given by $\mathcal{S}^{FF}_0\left[b\right]\left(\boldsymbol{r}\right) = \int_\Gamma b\left(\boldsymbol{r'}\right)$ which is zero if $\boldsymbol{b}$ is in the loop space. 
With a similar strategy, it can be derived that the electric far-field is given by
\begin{equation}
\boldsymbol{E}^{FF} \vcentcolon=\jmath\omega\boldsymbol{1}_r\times\left(\boldsymbol{1}_r\times \mathcal{S}^{FF}\left[\boldsymbol{v}\right]\right).
\end{equation}
\subsubsection{Near-Field}
\begin{figure}[h!]
\centering
\begin{subfigure}{0.48\columnwidth}
    \centering
    \includegraphics[width=\linewidth]{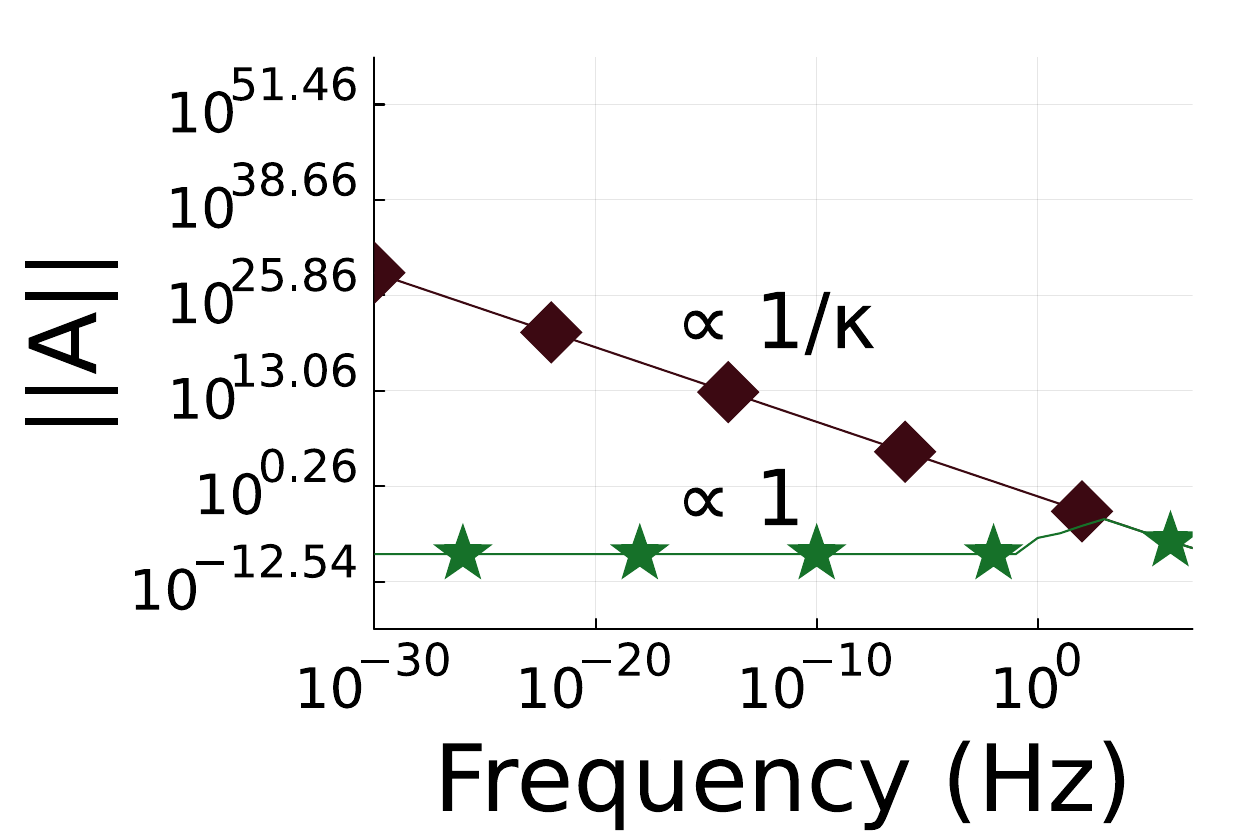}
    \caption{Vector-potential in the near field for (LF)-VPIE-C methods}
\end{subfigure}
\hfill
\begin{subfigure}{0.48\columnwidth}
    \centering
    \includegraphics[width=\linewidth]{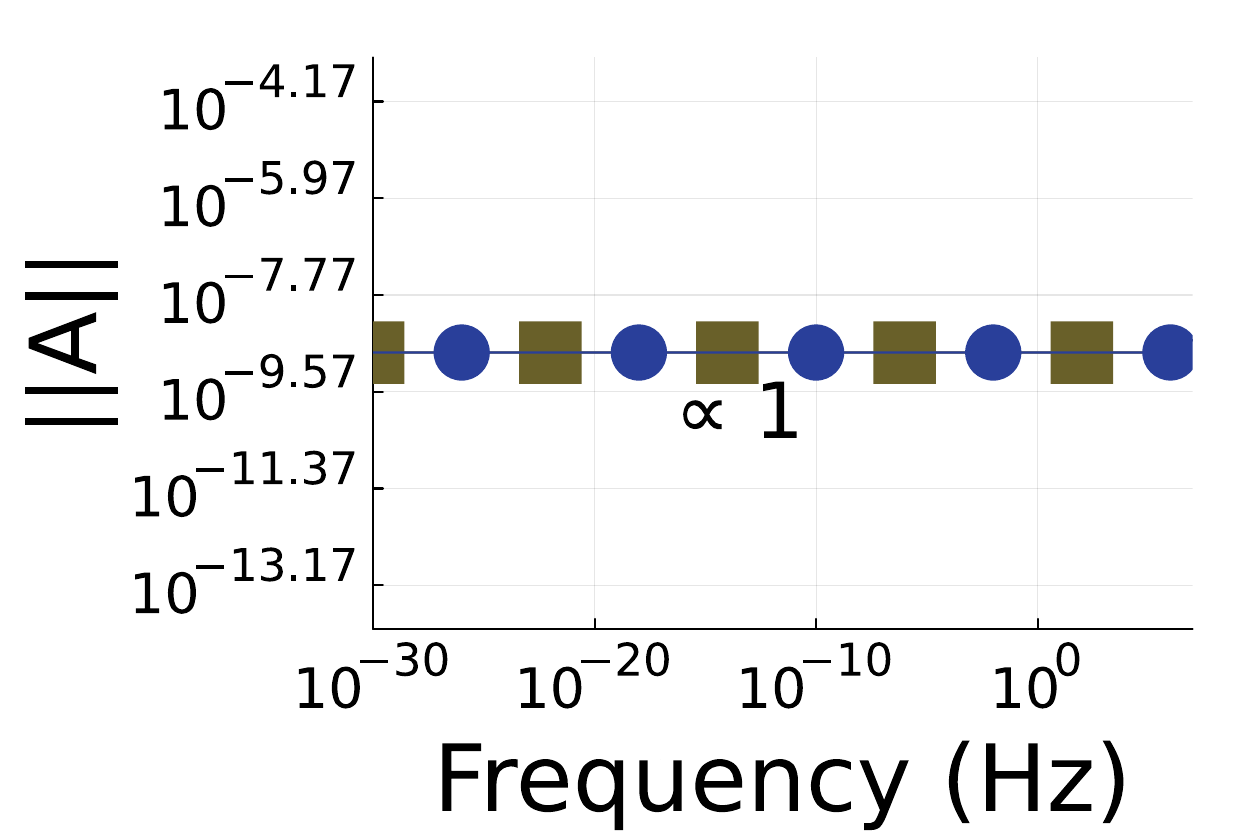}
    \caption{Vector-potential in the near field for (LF)-VPIE-V methods}
\end{subfigure}

\vspace{0.5em}
\begin{subfigure}{0.48\columnwidth}
    \centering
    \includegraphics[width=\linewidth]{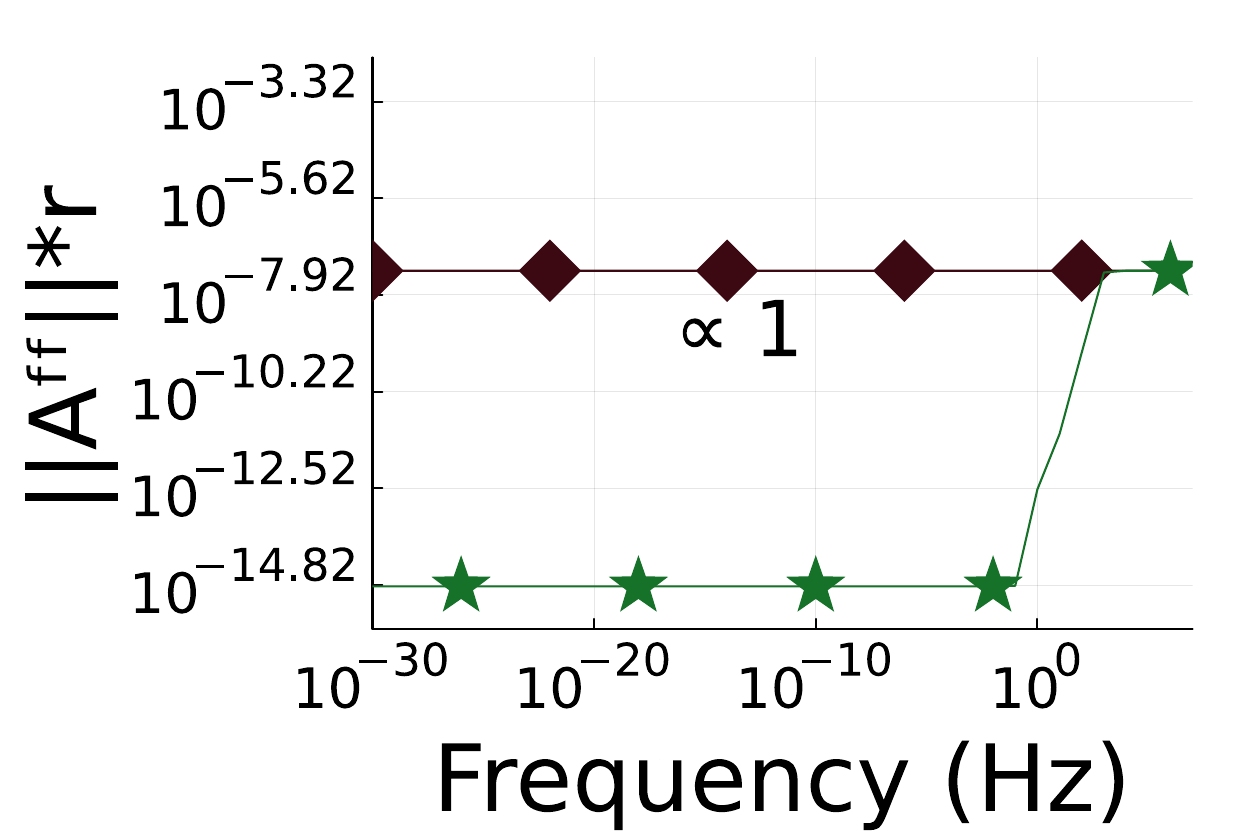}
    \caption{Vector-potential in the far field for (LF)-VPIE-C methods}
\end{subfigure}
\hfill
\begin{subfigure}{0.48\columnwidth}
    \centering
    \includegraphics[width=\linewidth]{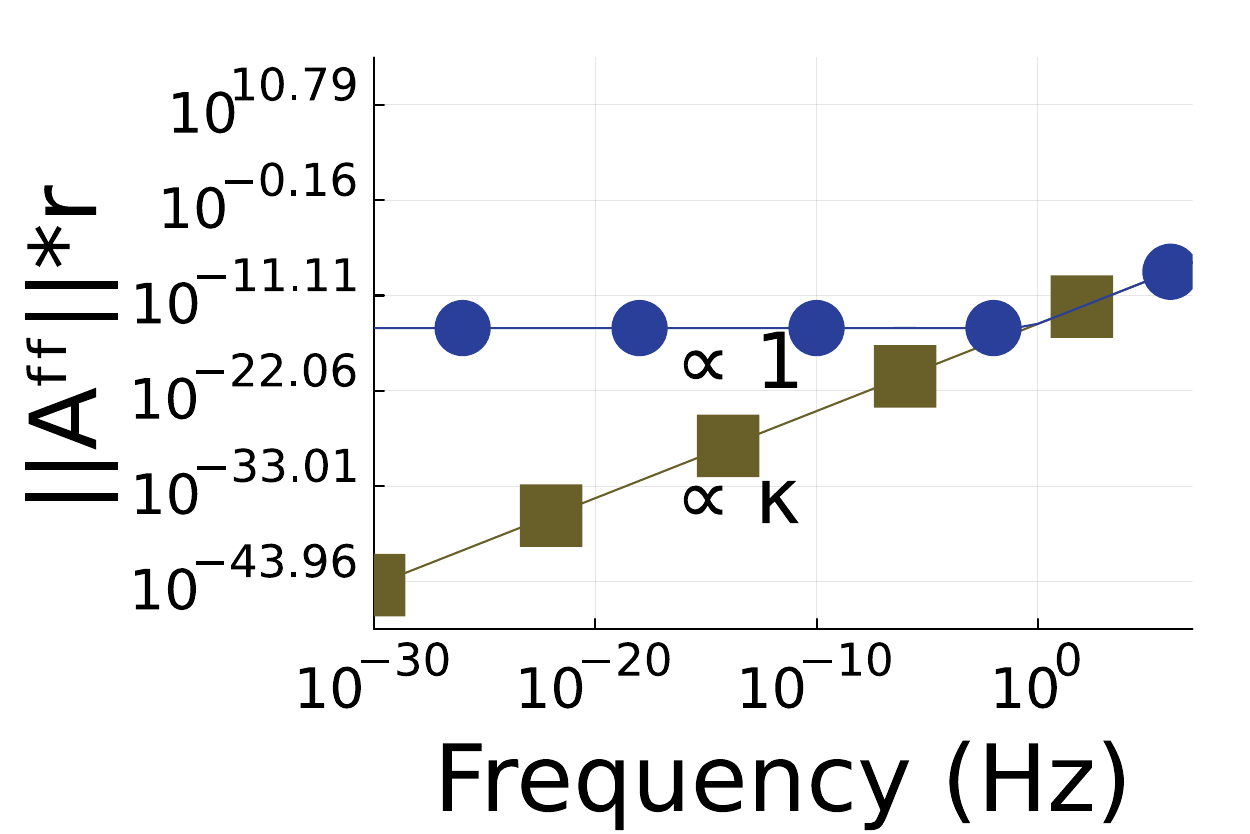}
    \caption{Vector-potential in the far field for (LF)-VPIE-V methods}
\end{subfigure}

\vspace{0.5em}

\begin{subfigure}{0.48\columnwidth}
    \centering
    \includegraphics[width=\linewidth]{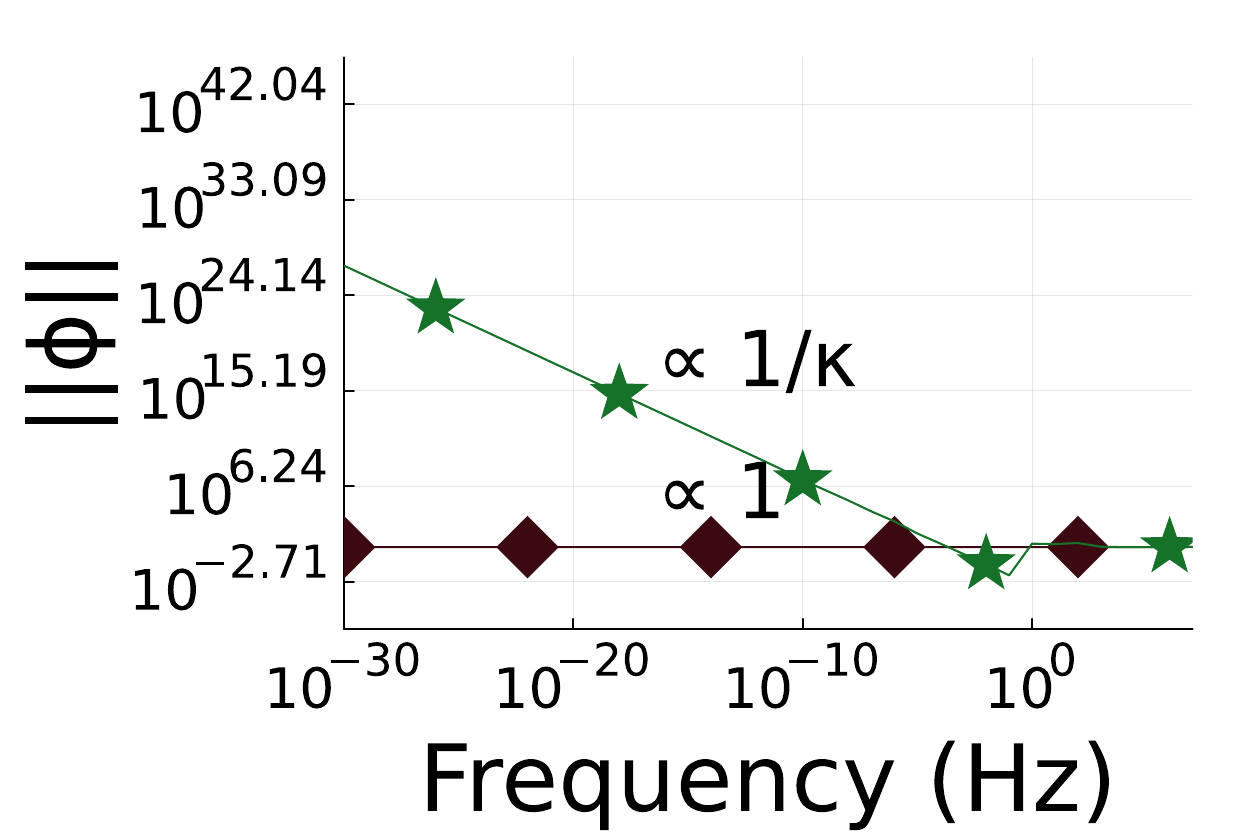}
    \caption{Scalar potential in the near field for (LF)-VPIE-C methods}
\end{subfigure}
\hfill
\begin{subfigure}{0.48\columnwidth}
    \centering
    \includegraphics[width=\linewidth]{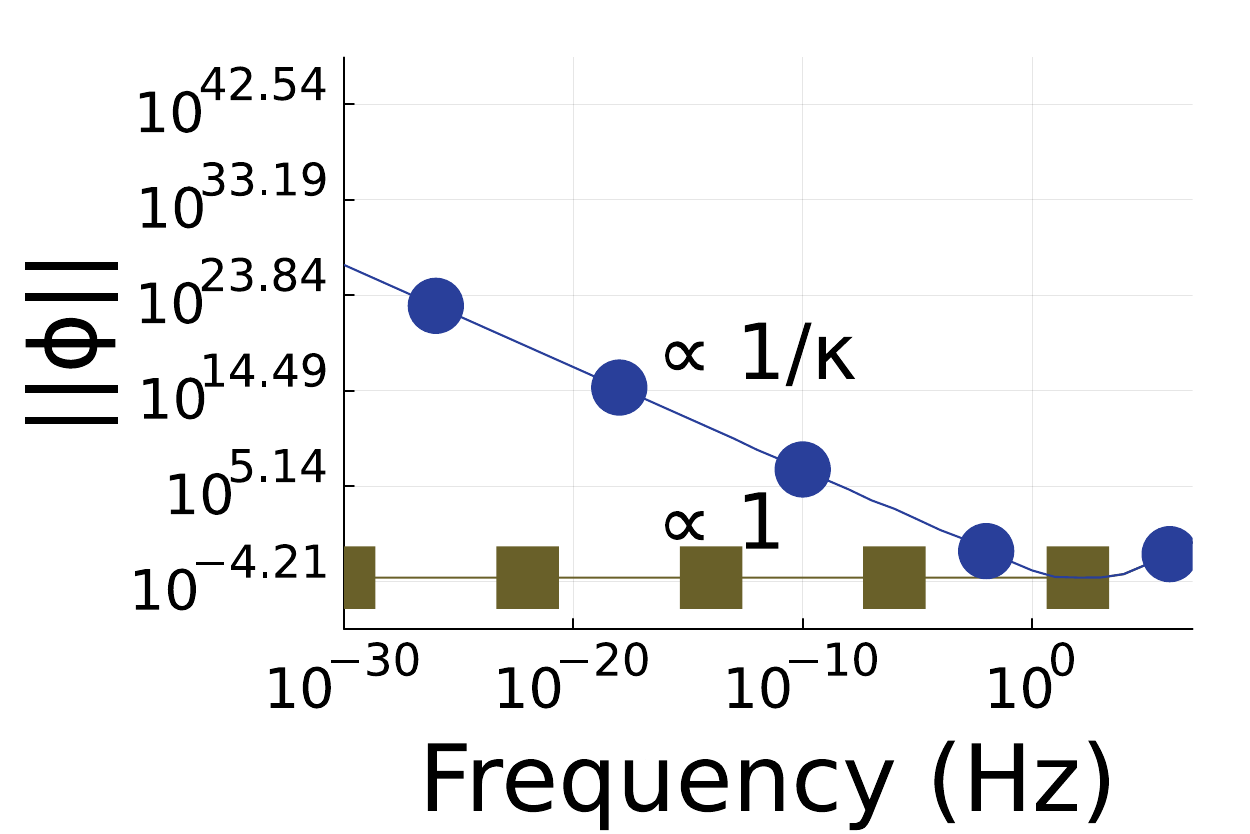}
    \caption{Scalar potential in the near field for (LF)-VPIE-V methods}
\end{subfigure}

\vspace{0.5em}

\begin{subfigure}{0.48\columnwidth}
    \centering
    \includegraphics[width=\linewidth]{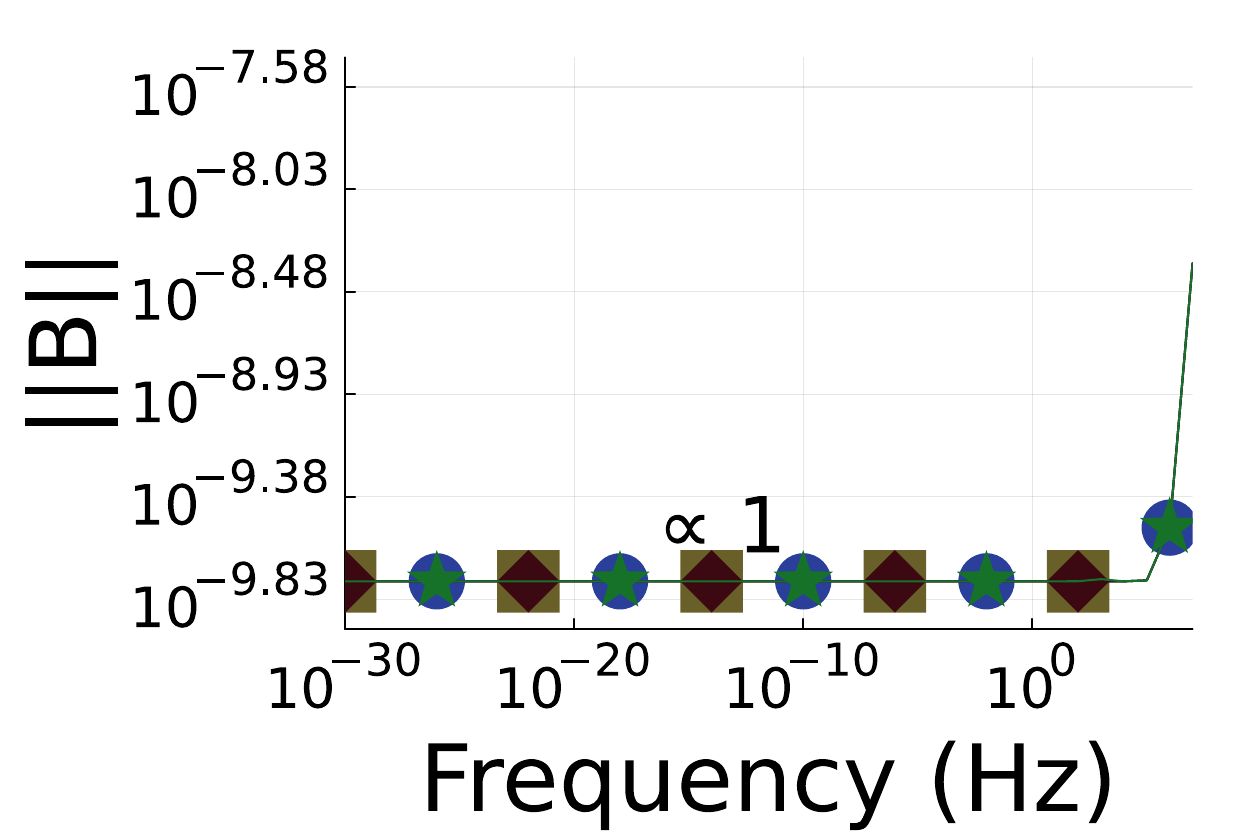}
    \caption{Magnetic field in the near field}
\end{subfigure}
\hfill
\begin{subfigure}{0.48\columnwidth}
    \centering
    \includegraphics[width=\linewidth]{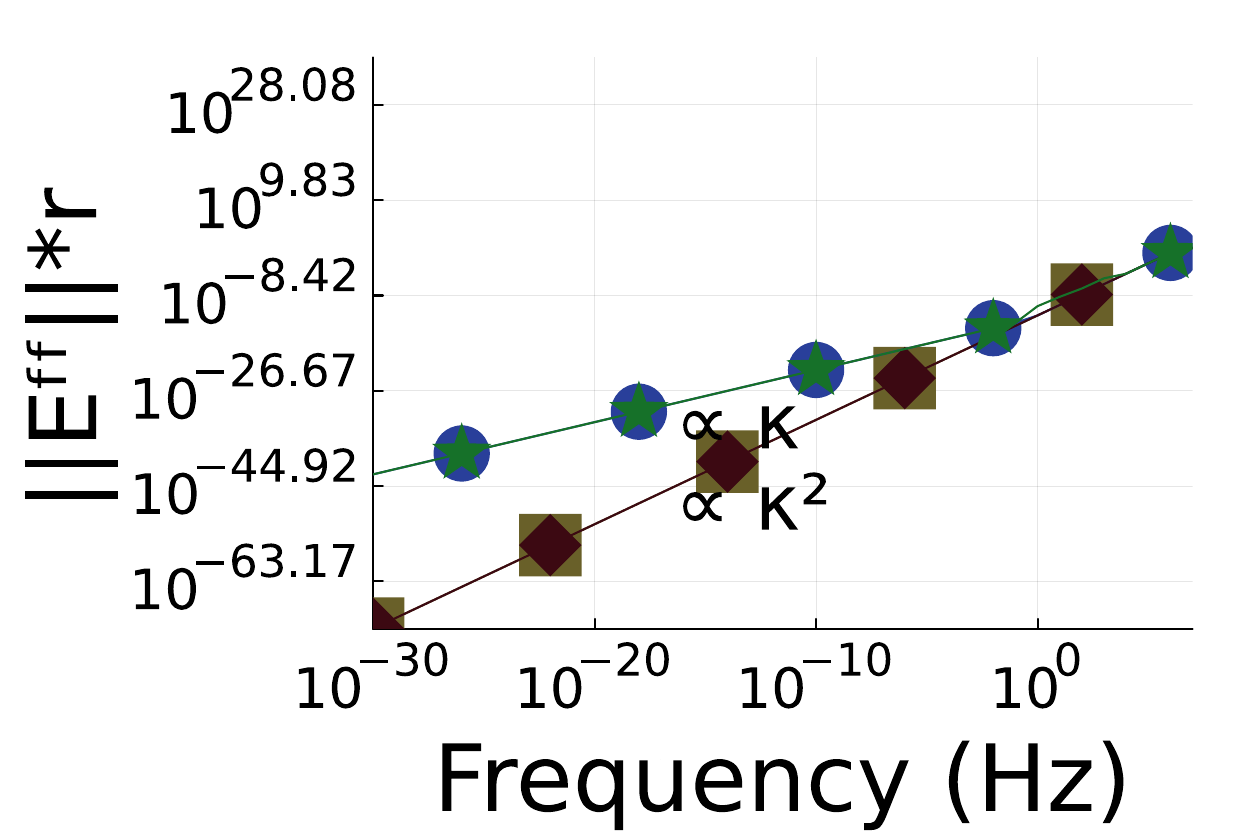}
    \caption{Electric field in the far field}
\end{subfigure}

\vspace{0.5em}

\begin{subfigure}{0.48\columnwidth}
    \centering
    \includegraphics[width=\linewidth]{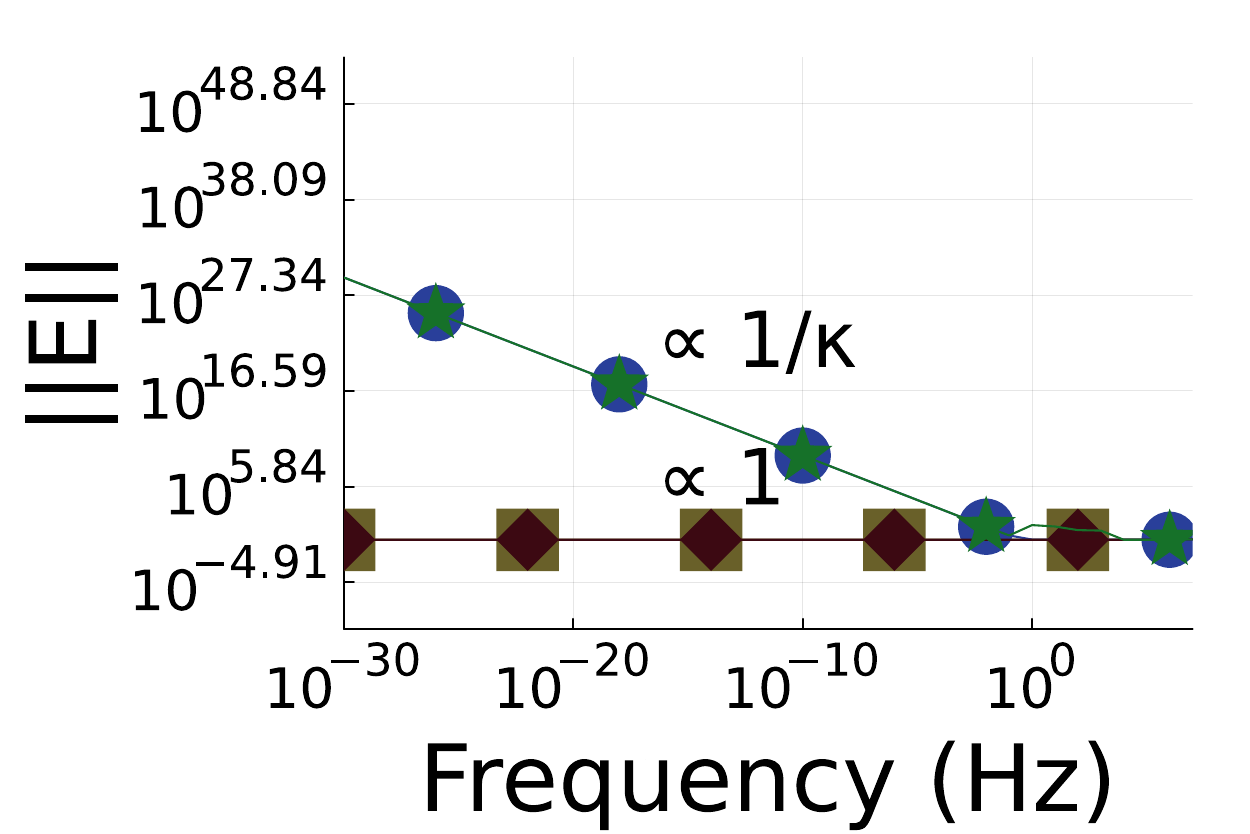}
    \caption{Electric field in the near field}
\end{subfigure}
\begin{subfigure}{\columnwidth}
\centering
\includegraphics[width=0.7\linewidth]{legend-1.pdf}
\end{subfigure}
\vspace{1em}
\caption{The scaling of the physical quantities with respect to the frequency for the (LF)-VPIE-(C/V) methods. Meshing parameter $h=0.2$. The near field is evaluated at $(4.0,4.0,2.0)$, the far field in the $\boldsymbol{1}_z$ direction.}
\label{fig:physical_quant}
\end{figure}

\begin{figure}
    \centering
    \includegraphics[width=\linewidth]{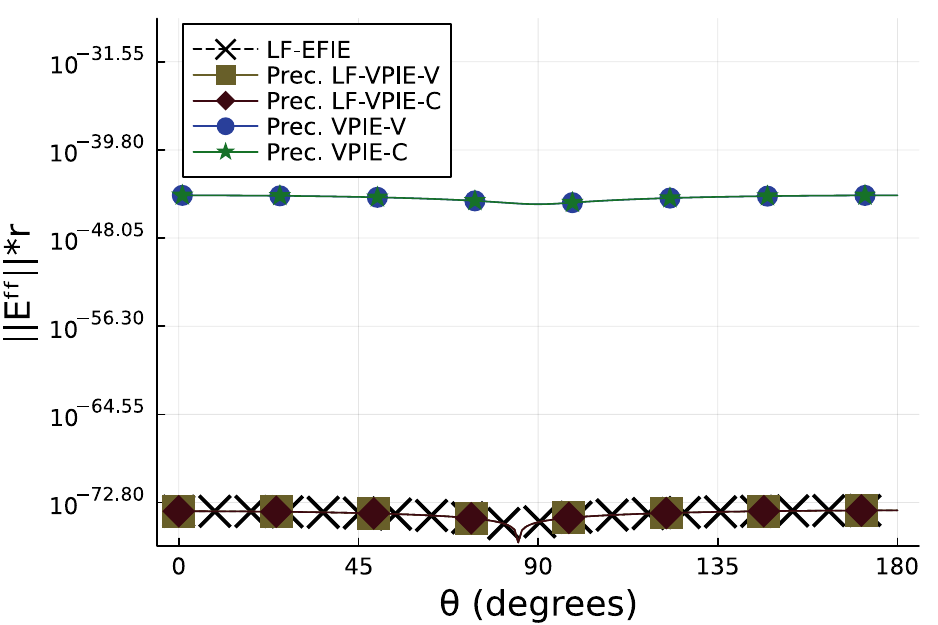}
    \caption{Comparison of the electric field evaluated in the far field and computed with the different methods against the low-frequency stable EFIE method from \cite{andriulli_well-conditioned_2013}, $\phi = 0.0$, $f=10^{-30}\text{Hz}$, $h=0.2$.}
    \label{fig:farfield}
\end{figure}

The scattered near-field of the scalar potential $\phi^{sc}$, the vector-potential $\boldsymbol{A}^{sc}$, the electric field $\boldsymbol{E}^{sc}$ and the magnetic field $\boldsymbol{B}^{sc}$ are studied. These quantities can be computed from \eqref{eq:Asc} and are expressed as
\begin{align}
    \phi^{sc} &\vcentcolon= \frac{j}{\omega\epsilon_0\mu_0}\left(\nabla\cdot\mathcal{S}_{\kappa}\left[\boldsymbol{v}\right]+\kappa^2\mathcal{S}_{\kappa}\left[w\right]-\nabla\cdot \mathcal{S}_{\kappa}\left[\boldsymbol{n}V_{\boldsymbol{A}}\right]\right)\\
    \boldsymbol{A}^{sc} &\vcentcolon= \mathcal{S}_{\kappa}\left[\boldsymbol{v}\right] - \nabla\mathcal{S}_{\kappa}\left[w\right] - \mathcal{S}_{\kappa}\left[\boldsymbol{n}V_{\boldsymbol{A}}\right]\\
    \boldsymbol{B}^{sc} &\vcentcolon= \nabla\times\mathcal{S}_{\kappa}\left[\boldsymbol{v}\right]\\
    \boldsymbol{E}^{sc} &\vcentcolon= - \frac{j}{\omega\epsilon_0\mu_0}\nabla\times\left(\nabla\times\mathcal{S}_{\kappa}\left[\boldsymbol{v}\right]\right).
\end{align}
The contribution of the different degrees of freedom for the different methods to those physical quantities can be studied in terms of their order of scaling in $\kappa$. The scaling of the different terms is summarized in tables \ref{tab:phys_scaling_VPIE_C} and \ref{tab:phys_scaling_VPIE_V}. We can conclude from those tables that, if a DOF should contribute in lowest-order to the computation of the physical quantity, but it does not do so because of the wrong scaling, or opposite, if a DOF should not contribute in lowest-order, but it does so because of wrong scaling, then there will be a significant error in the physical quantity and one should use the proposed low-frequency stabilized methods. \\
From this reasoning, we expect that the VPIE-C method might predict the magnetic field in the near field correctly, as the error introduced by the $\text{P}_\Sigma \text{v}$ term is smaller than that of the $\text{P}_\Lambda \text{v}$ term. The VPIE-V method is performing a little better as it might also predict the near-field vector-potential correctly by the same reasoning. \\
The actual comparison of the scaling of the physical quantities for the (LF)-VPIE-(C/V) methods is shown in Figs. \ref{fig:physical_quant}. Note that all four methods are shown altogether in the same figure for physical quantities that are gauge invariant.\\
A comparison to the low-frequency-stable electric field integral equation method from \cite{andriulli_well-conditioned_2013} is made for the electric field evaluated in the far-field in Fig. \ref{fig:farfield}, at a frequency of $10^{-30}$ Hz. Only the low-frequency-stable methods yield a correct result.

\begin{table}
\centering
\renewcommand{\arraystretch}{1.6}
\begin{NiceTabular}{p{0.075\textwidth}*{4}{wc{0.050\textwidth}}}[hvlines]

VPIE-C & \Block{2-1}{$\text{P}_\Lambda\text{v}$ }& \Block{2-1}{$\text{P}_\Sigma\text{v}$} & \Block{2-1}{$\text{P}_e\text{w}$ }& \Block{2-1}{$\text{P}_o\text{w}$ }\\ LF-VPIE-C \\

\Block{2-1}{$\boldsymbol{A}^{FF}$}
& $\propto \kappa$ & $\propto 1$ & $\propto \kappa$ & $\propto \kappa$ \\
& $\propto \kappa$ & $\propto \kappa$ & $\propto 1$ & $\propto \kappa$ \\

\Block{2-1}{$\boldsymbol{E}^{FF}$}
& $\propto \kappa^2$ & $\propto \kappa$ & / & / \\
& $\propto \kappa^2$ & $\propto \kappa^2$ & / & / \\

\Block{2-1}{$\phi^{sc}$}
& / & $\propto \kappa^{-1}$ & $\propto \kappa$ & $\propto \kappa$ \\
& / & $\propto 1$ & $\propto 1$ & $\propto \kappa$ \\

\Block{2-1}{$\boldsymbol{A}^{sc}$}
& $\propto 1$ & $\propto 1$ & $\propto 1$ & $\propto 1$ \\
& $\propto 1$ & $\propto \kappa$ & $\propto \kappa^{-1}$ & $\propto 1$ \\

\Block{2-1}{$\boldsymbol{B}^{sc}$}
& $\propto 1$ & $\propto 1$ & / & / \\
& $\propto 1$ & $\propto \kappa$ & / & / \\

\Block{2-1}{$\boldsymbol{E}^{sc}$}
& $\propto \kappa$ & $\propto \kappa^{-1}$ & / & / \\
& $\propto \kappa$ & $\propto 1$ & / & / \\

\end{NiceTabular}

\caption{The low-frequency scaling of the different terms. First row: VPIE-C, second row: LF-VPIE-C.}
\label{tab:phys_scaling_VPIE_C}
\end{table}

\begin{table}
\centering
\renewcommand{\arraystretch}{1.6}
\begin{NiceTabular}{p{0.075\textwidth}*{4}{wc{0.050\textwidth}}}[hvlines]

VPIE-V & \Block{2-1}{$\text{P}_\Lambda\text{v}$ }& \Block{2-1}{$\text{P}_\Sigma\text{v}$} & \Block{2-1}{$\text{w}$ }& \Block{2-1}{$\text{V}_{\boldsymbol{A}}$ }\\ LF-VPIE-V \\

\Block{2-1}{$\boldsymbol{A}^{FF}$}
& $\propto \kappa$ & $\propto 1$ & $\propto \kappa$ & $\propto \kappa$ \\
& $\propto \kappa$ & $\propto \kappa$ & $\propto \kappa$ & $\propto \kappa^2$ \\

\Block{2-1}{$\boldsymbol{E}^{FF}$}
& $\propto \kappa^2$ & $\propto \kappa$ & / & / \\
& $\propto \kappa^2$ & $\propto \kappa^2$ & / & / \\

\Block{2-1}{$\phi^{sc}$}
& / & $\propto \kappa^{-1}$ & $\propto \kappa$ & $\propto \kappa$ \\
& / & $\propto 1$ & $\propto \kappa$ & $\propto \kappa^2$ \\

\Block{2-1}{$\boldsymbol{A}^{sc}$}
& $\propto 1$ & $\propto 1$ & $\propto 1$ & $\propto 1$ \\
& $\propto 1$ & $\propto \kappa$ & $\propto 1$ & $\propto \kappa$ \\

\Block{2-1}{$\boldsymbol{B}^{sc}$}
& $\propto 1$ & $\propto 1$ & / & / \\
& $\propto 1$ & $\propto \kappa$ & / & / \\

\Block{2-1}{$\boldsymbol{E}^{sc}$}
& $\propto \kappa$ & $\propto \kappa^{-1}$ & / & / \\
& $\propto \kappa$ & $\propto 1$ & / & / \\

\end{NiceTabular}

\caption{The low-frequency scaling of the different terms. First row: VPIE-V, second row: LF-VPIE-V.}
\label{tab:phys_scaling_VPIE_V}
\end{table}

\subsection{Calderón Preconditioner}
\begin{figure}
    \centering
    \includegraphics[width=\linewidth]{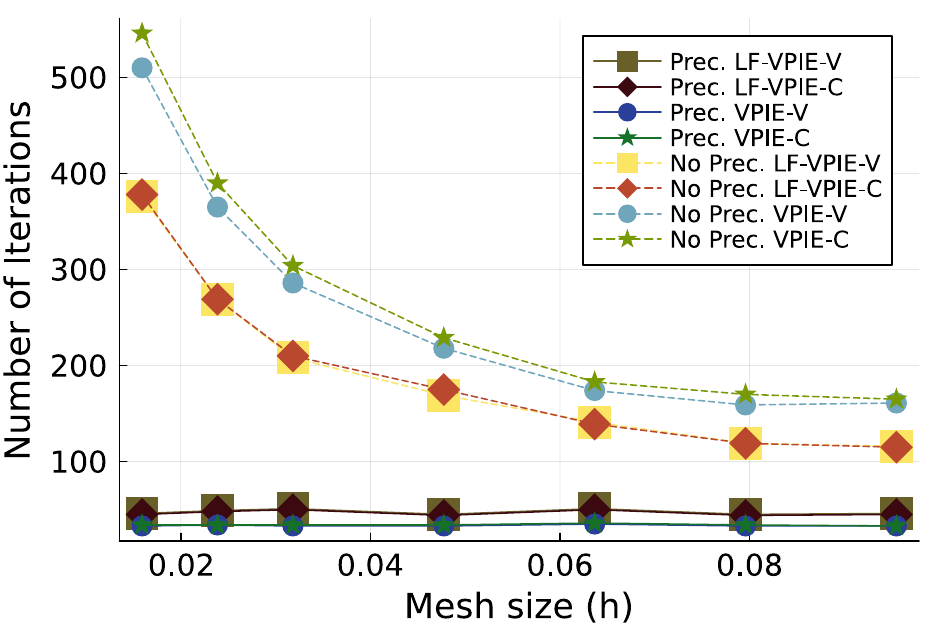}
    \caption{The number of iterations as a function of the meshingparameter $h$ at $f=10^{-30}\text{Hz}$.}
    \label{fig:it_h}
\end{figure}

The number of iterations as a function of the meshing parameter $h$, which is the mean edge-length in the conformal mesh, is shown in Fig. \ref{fig:it_h}. The number of iterations for the preconditioned system does not depend on $h$. The smallest mesh with $h=0.1$ contains 38408 triangles, which corresponds to 96020 degrees of freedom for the (LF)-VPIE-C methods, including both the scalar and vector ones.

\begin{figure}
    \centering
    \includegraphics[width=\linewidth]{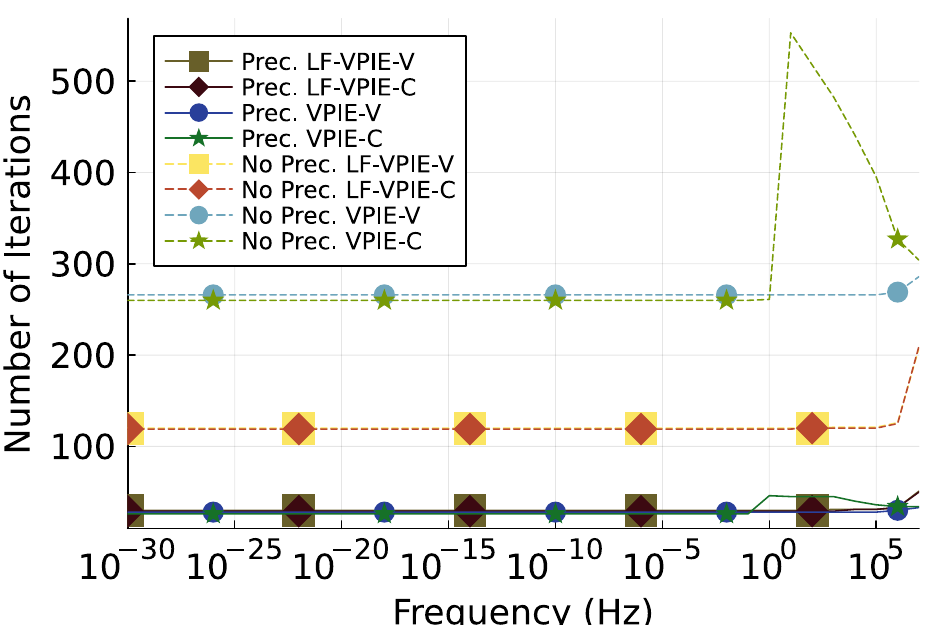}
    \caption{Number of iterations as a function of the frequency at $h=0.2$.}
    \label{fig:it_freq}
\end{figure}

The number of iterations with respect to the frequency at $h=0.2$, is shown in Fig. \ref{fig:it_freq} for both the preconditioned and not preconditioned systems. The number of iterations stays constant when the frequency approaches zero, but the total number of iterations for the preconditioned systems is lower than for the systems that are not preconditioned.

\section{Conclusion}
Two stabilized Calderón preconditioned vector-potential-only methods for modeling electromagnetic scattering at a perfect electric conductor (PEC) were proposed. The scalar potential is recovered in the post-processing with the Lorenz gauge. The stabilized LF-VPIE-C and LF-VPIE-V methods, based on quasi-Helmholtz projectors, demonstrate correct scaling as a function of the frequency, in both the degrees of freedom and in the resulting electromagnetic fields. Furthermore, the computed electric far field is in close agreement with the far field obtained from the stabilized electric field integral equation, used as a reference solution. The Calderón preconditioned methods do not suffer from the dense-mesh breakdown. The number of iterations of the iterative Krylov solver is independent of the meshing parameter $h$.
\section*{ACKNOWLEDGMENT}
This project has received funding from the European Research Council (ERC) under the European Union’s Horizon 2020 research and innovation programme (Grant agreement No. 101001847).

\bibliographystyle{ieeetr}
\bibliography{bibliografieapsaps2025}

@article{hawkins_analytic_2023,
	title = {Analytic {Preconditioners} for {Decoupled} {Potential} {Integral} {Equations} and {Wideband} {Analysis} of {Scattering} {From} {PEC} {Objects}},
	volume = {71},
	issn = {1558-2221},
	url = {https://ieeexplore.ieee.org/abstract/document/10153463},
	doi = {10.1109/TAP.2023.3284490},
	number = {8},
	urldate = {2024-02-13},
	journal = {IEEE Transactions on Antennas and Propagation},
	author = {Hawkins, Jacob A. and Baumann, Luke and Aktulga, H. M. and Dault, D. and Shanker, B.},
	month = aug,
	year = {2023},
	pages = {6753--6765},
}

@ARTICLE{GlobalMultiCaldPrec,
  author={Olyslager, Paul and Rogier, Hendrik and Cools, Kristof},
  journal={IEEE Transactions on Antennas and Propagation}, 
  title={Calderón Preconditioners for the Global Multi-Trace Vector Potential Integral Equation}, 
  year={2025},
  volume={73},
  number={11},
  pages={9167-9176},
  doi={10.1109/TAP.2025.3598520}}

@ARTICLE{VPIE_C_chew,
  author={Liu, Qin S. and Sun, Sheng and Chew, Weng Cho},
  journal={IEEE Transactions on Antennas and Propagation}, 
  title={A Potential-Based Integral Equation Method for Low-Frequency Electromagnetic Problems}, 
  year={2018},
  volume={66},
  number={3},
  pages={1413-1426},
  doi={10.1109/TAP.2018.2794388}}

@article{vico_decoupled_2016,
	title = {The {Decoupled} {Potential} {Integral} {Equation} for {Time}-{Harmonic} {Electromagnetic} {Scattering}},
	volume = {69},
	copyright = {© 2016 Wiley Periodicals, Inc.},
	issn = {1097-0312},
	url = {https://onlinelibrary.wiley.com/doi/abs/10.1002/cpa.21585},
	doi = {10.1002/cpa.21585},
	language = {en},
	number = {4},
	urldate = {2024-02-13},
	journal = {Communications on Pure and Applied Mathematics},
	author = {Vico, Felipe and Ferrando, Miguel and Greengard, Leslie and Gimbutas, Zydrunas},
	year = {2016},
	pages = {771--812},
}

@article{schulz_coupled_2022,
	title = {Coupled {Domain}-{Boundary} {Variational} {Formulations} for {Hodge}–{Helmholtz} {Operators}},
	volume = {94},
	issn = {1420-8989},
	url = {https://doi.org/10.1007/s00020-022-02684-6},
	doi = {10.1007/s00020-022-02684-6},
	language = {en},
	number = {1},
	urldate = {2024-02-13},
	journal = {Integral Equations and Operator Theory},
	author = {Schulz, Erick and Hiptmair, Ralf},
	month = feb,
	year = {2022},
	pages = {7},
}

@article{HIPTMAIR2006699,
title = {Operator Preconditioning},
journal = {Computers \& Mathematics with Applications},
volume = {52},
number = {5},
pages = {699-706},
year = {2006},
note = {Hot Topics in Applied and Industrial Mathematics},
issn = {0898-1221},
doi = {https://doi.org/10.1016/j.camwa.2006.10.008},
url = {https://www.sciencedirect.com/science/article/pii/S0898122106002495},
author = {R. Hiptmair}
}

@article{rao_electromagnetic_1982,
	title = {Electromagnetic scattering by surfaces of arbitrary shape},
	volume = {30},
	copyright = {https://ieeexplore.ieee.org/Xplorehelp/downloads/license-information/IEEE.html},
	issn = {0018-926X, 1558-2221},
	url = {https://ieeexplore.ieee.org/document/1142818/},
	doi = {10.1109/TAP.1982.1142818},
	language = {en},
	number = {3},
	urldate = {2024-06-27},
	journal = {IEEE Transactions on Antennas and Propagation},
	author = {Rao, S. and Wilton, D. and Glisson, A.},
	month = may,
	year = {1982},
	pages = {409--418},
}

@article{chen_low-frequency_2022,
	title = {On the {Low}-{Frequency} {Behavior} of {Vector} {Potential} {Integral} {Equations} for {Perfect} {Electrically} {Conducting} {Scatterers}},
	volume = {70},
	issn = {1558-2221},
	url = {https://ieeexplore.ieee.org/document/9912315/references#references},
	doi = {10.1109/TAP.2022.3210650},
	number = {12},
	urldate = {2024-12-12},
	journal = {IEEE Transactions on Antennas and Propagation},
	author = {Chen, Rui and Ulku, H. Arda and Andriulli, Francesco P. and Bagci, Hakan},
	month = dec,
	year = {2022},
	pages = {12411--12416},
}

@article{buffa_dual_2005,
	title = {A dual finite element complex on the barycentric refinement},
	volume = {340},
	issn = {1631-073X},
	url = {https://www.sciencedirect.com/science/article/pii/S1631073X04006041},
	doi = {10.1016/j.crma.2004.12.022},
	number = {6},
	urldate = {2024-04-11},
	journal = {Comptes Rendus Mathematique},
	author = {Buffa, Annalisa and Christiansen, Snorre H.},
	month = mar,
	year = {2005},
	pages = {461--464},
}

@ARTICLE{Excitation-Aware-Bernd,
  author={Hofmann, Bernd and Eibert, Thomas F. and Andriulli, Francesco P. and Adrian, Simon B.},
  journal={IEEE Transactions on Antennas and Propagation}, 
  title={An Excitation-Aware and Self-Adaptive Frequency Normalization for Low-Frequency Stabilized Electric Field Integral Equation Formulations}, 
  year={2023},
  volume={71},
  number={5},
  pages={4301-4314},
  doi={10.1109/TAP.2023.3247896}}

@ARTICLE{andriulli_cond_loop_star,
  author={Andriulli, Francesco P.},
  journal={IEEE Transactions on Antennas and Propagation}, 
  title={Loop-Star and Loop-Tree Decompositions: Analysis and Efficient Algorithms}, 
  year={2012},
  volume={60},
  number={5},
  pages={2347-2356},
  doi={10.1109/TAP.2012.2189723}}

@article{andriulli_well-conditioned_2013,
	title = {On a {Well}-{Conditioned} {Electric} {Field} {Integral} {Operator} for {Multiply} {Connected} {Geometries}},
	volume = {61},
	issn = {1558-2221},
	url = {https://ieeexplore.ieee.org/document/6381461},
	doi = {10.1109/TAP.2012.2234072},
	number = {4},
	urldate = {2024-12-12},
	journal = {IEEE Transactions on Antennas and Propagation},
	author = {Andriulli, Francesco P. and Cools, Kristof and Bogaert, Ignace and Michielssen, Eric},
	month = apr,
	year = {2013},
	pages = {2077--2087},
}

@article{sharma_electromagnetic_2022,
	title = {Electromagnetic {Modeling} of {Lossy} {Interconnects} {From} {DC} to {High} {Frequencies} {With} a {Potential}-{Based} {Boundary} {Element} {Formulation}},
	volume = {70},
	issn = {1557-9670},
	url = {https://ieeexplore.ieee.org/abstract/document/9802859/figures#figures},
	doi = {10.1109/TMTT.2022.3180390},
	number = {8},
	urldate = {2024-12-13},
	journal = {IEEE Transactions on Microwave Theory and Techniques},
	author = {Sharma, Shashwat and Triverio, Piero},
	month = aug,
	year = {2022},
	pages = {3847--3861},
}

@INPROCEEDINGS{Triv_prec_los_cond_pot,
  author={Sharma, Shashwat and Triverio, Piero},
  booktitle={2022 IEEE International Symposium on Antennas and Propagation and USNC-URSI Radio Science Meeting (AP-S/URSI)}, 
  title={Preconditioned Potential-Based Surface Integral Method for Modeling Lossy Conductors From DC to High Frequencies}, 
  year={2022},
  volume={},
  number={},
  pages={503-504},
  doi={10.1109/AP-S/USNC-URSI47032.2022.9886874}}

@article{low_rank_prec,
title = {On the {Nyström} discretization of integral equations on planar curves with corners},
journal = {Applied and Computational Harmonic Analysis},
volume = {32},
number = {1},
pages = {45-64},
year = {2012},
issn = {1063-5203},
doi = {https://doi.org/10.1016/j.acha.2011.03.002},
url = {https://www.sciencedirect.com/science/article/pii/S1063520311000297},
author = {James Bremer}
}

@inproceedings{gur_low-frequency_2017,
	title = {Low-frequency breakdown of the potential integral equations and its remedy},
	url = {https://ieeexplore.ieee.org/document/8293221},
	doi = {10.1109/PIERS-FALL.2017.8293221},
	urldate = {2024-12-13},
	booktitle = {2017 {Progress} in {Electromagnetics} {Research} {Symposium} - {Fall} ({PIERS} - {FALL})},
	author = {Gür, U. M. and Ergül, O.},
	month = nov,
	year = {2017},
	pages = {676--682},
}

@article{EMQMFinite_element,
title = {A leap-frog finite element method for wave propagation of Maxwell–Schrödinger equations with nonlocal effect in metamaterials},
journal = {Computers \& Mathematics with Applications},
volume = {90},
pages = {25-37},
year = {2021},
issn = {0898-1221},
doi = {https://doi.org/10.1016/j.camwa.2021.02.019},
url = {https://www.sciencedirect.com/science/article/pii/S0898122121000675},
author = {C.H. Yao and Z.Y. Wang and Y.M. Zhao}
}

@ARTICLE{chienlftd,
  author={Le, Van Chien and Cordel, Pierrick and Andriulli, Francesco P. and Cools, Kristof},
  journal={IEEE Transactions on Antennas and Propagation}, 
  title={A Stabilized Time-Domain Combined Field Integral Equation Using the Quasi-Helmholtz Projectors}, 
  year={2024},
  volume={72},
  number={7},
  pages={5852-5864},
  doi={10.1109/TAP.2024.3410709}}

@ARTICLE{EMQMFDTD,
  author={Decleer, Pieter and Vande Ginste, Dries},
  journal={IEEE Journal on Multiscale and Multiphysics Computational Techniques}, 
  title={A Hybrid {EM/QM} Framework Based on the {ADHIE-FDTD} Method for the Modeling of Nanowires}, 
  year={2022},
  volume={7},
  number={},
  pages={236-251},
  doi={10.1109/JMMCT.2022.3198750}}

@ARTICLE{TD-FD_relation,

  author={Beghein, Yves and Cools, Kristof and Andriulli, Francesco P.},

  journal={IEEE Transactions on Antennas and Propagation}, 

  title={A DC-Stable, Well-Balanced, Calderón Preconditioned Time Domain Electric Field Integral Equation}, 

  year={2015},

  volume={63},

  number={12},

  pages={5650-5660},

  doi={10.1109/TAP.2015.2487500}}

@ARTICLE{QHP_VS_LS,

  author={Adrian, Simon B. and Dély, Alexandre and Consoli, Davide and Merlini, Adrien and Andriulli, Francesco P.},

  journal={IEEE Open Journal of Antennas and Propagation}, 

  title={Electromagnetic Integral Equations: Insights in Conditioning and Preconditioning}, 

  year={2021},

  volume={2},

  number={},

  pages={1143-1174},

  doi={10.1109/OJAP.2021.3121097}}

@article{claeys_first-kind_nodate,
	title = {First-{Kind} {Boundary} {Integral} {Equations} for the {Hodge}-{Helmholtz} {Equation}},
	language = {en},
	author = {Claeys, X and Hiptmair, R},
}

@ARTICLE{MFIE_LOW_FREQ,

  author={Merlini, Adrien and Beghein, Yves and Cools, Kristof and Michielssen, Eric and Andriulli, Francesco P.},

  journal={IEEE Transactions on Antennas and Propagation}, 

  title={Magnetic and Combined Field Integral Equations Based on the Quasi-Helmholtz Projectors}, 

  year={2020},

  volume={68},

  number={5},

  pages={3834-3846},

  doi={10.1109/TAP.2020.2964941}}

@ARTICLE{MFIE_Low_Freq_2,

  author={Bogaert, Ignace and Cools, Kristof and Andriulli, Francesco P. and Bağcı, Hakan},

  journal={IEEE Transactions on Antennas and Propagation}, 

  title={Low-Frequency Scaling of the Standard and Mixed Magnetic Field and Müller Integral Equations}, 

  year={2014},

  volume={62},

  number={2},

  pages={822-831},

  doi={10.1109/TAP.2013.2293783}}

\end{document}